\documentclass[twocolumn,superscriptaddress,amsmath,amssymb,longbibliography]{revtex4-1}
\usepackage[T1]{fontenc} 
\usepackage[utf8]{inputenc}
\usepackage{multirow}
\usepackage{float}
\usepackage{bm}
\usepackage{booktabs} 
\usepackage{array} 
\usepackage{graphicx}
\usepackage{amsmath}
\usepackage{amssymb}
\usepackage{color}
\usepackage{ulem}
\usepackage[per-mode=symbol,separate-uncertainty]{siunitx} 
\usepackage{natbib}
\usepackage{physics}

\newcommand{\parallelsum}{\mathbin{\!/\mkern-5mu/\!}}

\renewcommand{\thetable}{\arabic{table}}

\begin{document}

\author{S\'ebastien L\'eger} 
\author{Javier Puertas-Mart\'inez}
\author{Karthik Bharadwaj}
\author{R\'emy Dassonneville}
\author{Jovian Delaforce}
\author{Farshad Foroughi} 
\author{Vladimir Milchakov}
\author{Luca Planat}
\author{Olivier Buisson} 
\author{C\'ecile Naud}
\author{Wiebke Hasch-Guichard} 
\author{Serge Florens}
\affiliation{Univ. Grenoble Alpes, CNRS, Grenoble INP, Institut N\'eel, 38000 Grenoble, France}
\author{Izak Snyman} 
\affiliation{Mandelstam Institute for Theoretical Physics, School of Physics, University of the Witwatersrand, 
Johannesburg, South Africa}
\author{Nicolas Roch}
\email[]{Correspondence and requests for materials should be addressed to N.R. (email: nicolas.roch@neel.cnrs.fr)}
\affiliation{Univ. Grenoble Alpes, CNRS, Grenoble INP, Institut N\'eel, 38000 Grenoble, France}

\title{Observation of quantum many-body effects due to zero point fluctuations in superconducting circuits}

\maketitle

\noindent
{\bf Electromagnetic fields  possess zero point fluctuations (ZPF) which lead to observable 
effects such as the Lamb shift and the Casimir effect. In the traditional quantum optics domain, these corrections remain perturbative due to the smallness of the fine structure constant. To provide a direct observation of non-perturbative effects driven by  ZPF in an open quantum system we  wire a highly non-linear Josephson junction to a high impedance transmission line, allowing large phase fluctuations across the junction. Consequently, the resonance of the former acquires a relative frequency shift that is orders of magnitude larger than for natural atoms. Detailed modelling  confirms that this renormalization is non-linear and quantum. Remarkably, the junction transfers its non-linearity to about 30 environmental modes, a striking back-action effect that transcends the standard Caldeira-Leggett paradigm. This work opens many exciting prospects for longstanding quests such as the tailoring of many-body Hamiltonians in the strongly non-linear regime, the observation of Bloch oscillations, or the development of high-impedance qubits.}

\noindent{\large \textbf{Introduction} \par}
The realization of many-body effects in quantum matter, often associated with
remarkable physical properties, hinges on strong interactions between constituents.
Mechanisms to achieve strong interactions include the Coulomb interaction in narrow band electronic
materials, and Feshbach resonances that can produce arbitrarily large scattering lengths in ultra-cold atomic gases. In contrast, while providing great design versatility, purely photonic
platforms,~\cite{Greentree:2006jg,Carusotto:2013gh,LeHur:2016ita} are not easily amenable to realizing strong correlations, since they usually come with weak non-linearity.
To circumvent this, superconducting circuits, which operate
in the microwave range and display high tunability, have been proposed~\cite{Houck:2012iq} for the 
exploration of correlated states of light.
Here, correlations originate from non-linear elements, such as Josephson junctions, and the 
enhancement of non-linearities is accompanied by large zero point fluctuations (ZPF).
This can be understood in the electronics language of impedance as follows.
The dynamics of a Josephson junction
is described by two conjugate variables: the number of transferred Cooper
pairs $\hat{n}$ and the superconducting phase difference $\hat{\phi}$.
Despite being an anharmonic oscillator, a Josephson junction with Josephson energy
$E_\text{J}$ and charging energy $E_\text{c}$, can be associated to an
impedance $Z_\text{J}=\hbar/(2e)^2 \sqrt{2 E_\text{c}/E_\text{J}}$ which sets
the amplitude of the fluctuations of $\hat{n}$ and $\hat{\phi}$. When $\big<\hat{\phi}^2\big>$ is sufficiently smaller than  unity,
$\big<\hat{\phi}^2\big> \propto Z_\text{J}/R_\text{Q}$ and $\big<\hat{n}^2\big>
\propto R_\text{Q}/Z_\text{J}$, with $R_\text{Q}=h/(2e)^2\simeq\SI{6.5}{\kilo\ohm}$
the superconducting quantum of resistance. Consequently, at low $Z_\text{J}$,
phase fluctuations are weak and the anharmonic Josephson cosine potential $E_\text{J}(1 - \cos \hat{\phi})$ can be
reduced to a quadratic potential plus a quartic perturbation, as is the case for
the transmon qubit~\cite{Koch:2007gz}. On the other hand, if $Z_\text{J}$ is large, the
full cosine potential is explored due to strong phase fluctuations.
Anharmonicity then becomes important, as observed with the Cooper-pair
box~\cite{Vion:2002uo} or the fluxonium qubit~~\cite{Manucharyan:2009fo}, and as
a result, the oscillation frequency $\omega_\text{J}$ can strongly deviate from the
harmonic value $\sqrt{2 E_\text{J} E_\text{C}}$. Thus, exploring many-body
physics in circuit quantum electrodynamics must rely on a careful tailoring of
ZPF.

The approach~\cite{Weiss} that we follow to explore many-body effects originating  
from a single non-linear superconducting element is to couple it to many harmonic modes.
In the presence of such an environment, the degree of anharmonicity of a non-linear
junction will also depend on the external impedance, and three regimes can be
identified. When $Z_\text{J}$ does not match the environmental impedance
$Z_\text{env}(\omega)$ at frequencies $\omega$ close to $\omega_\text{J}$, the
junction is accurately described as an almost isolated system, so that the
effect of the environment only amounts to small perturbative corrections,
similar to the Lamb shift~\cite{Fragner:2008ct}. At the same time, an
impedance-mismatched environment remains weakly perturbed by the non-linear
junction, and this absence of back-action allows it to be described as a set of
harmonic oscillators, following the Caldeira-Leggett
approach~\cite{Leggett:1987r_}. This simplified description is at the core of
the current understanding of open quantum systems, and was already verified
experimentally in the early studies of macroscopic quantum
tunneling~\cite{clarke_quantum-mechanics_1988}.
The important role of ZPF in the damping effect that such an environment has on a Josephson junction
was already noticed experimentally~\cite{Schwartz:1985} and explained theoretically~\cite{Zaikin:1986,Panyukov:1988} three 
decades ago. In these early works, the effect of ZPF was to renormalize junction properties such as the
critical DC current by about $1\%$, which nonetheless had a large effect on macroscopic quantum tunnelling rates.
When $Z_\text{env}\sim Z_\text{J}\ll R_\text{Q}$, the junctions and its
environment fully hybridize, since they are impedance matched, but the
anharmonicity of the junction remains weak and can be treated
pertubatively~\cite{Nigg:2012ek,Bourassa2012,Weissl:2015do,PuertasMartinez:2019gk}. 
The case $ Z_\text{env}\sim
Z_\text{J} \sim R_\text{Q}$ is much more challenging, both experimentally and
theoretically since the strongly anharmonic junction hybridises 
with many modes of its environment. In DC measurements, such effects result in the celebrated 
Schmid-Bulgadaev transition predicted more than thirty years ago~\cite{Schmid:1983vi,Lett:wq}, a 
localization phenomenon whose relevance for microwave AC measurements requires further experimental and theoretical investigations~\cite{Murani:2019vw}.
The environment provides a strong action on the junction,
which itself induces a sizeable back-action on many modes of the environment, the combined circuit
forming a complex many-body system reminiscent of quantum impurity problems
encountered in condensed matter~\cite{Schon:1990kj}. More specifically, the frequency
shift induced by the environment on the junction can be comparable to
$\omega_\text{J}$, a non-perturbative effect due to
a modification of the vacuum~\cite{Hekking:1997fs}.
At the same time, the non-linearity of the junction is transferred into the environmental
modes, affecting for instance their broadening, and producing a physical regime that was not addressed so far.

In this work, we report on the effects of zero point fluctuations in a device consisting of
a fully characterized multi-mode environment and a highly non-linear single Josephson junction,
acting as a weak link between two linear transmission lines,
with all subsystems reaching the high impedance regime. As a result, the transmission
of single photons through our device is strongly affected by the interplay of
non-linearities and zero point fluctuations.
We observe a 30\% renormalization of the junction frequency as compared to the value that would
have been obtained without ZPF -- analogous to a giant Lamb shift -- and we provide clear evidence
for modifications of the environmental vacuum, which inherits strong non-linear effects.
A detailed temperature analysis of our system proves the quantum origin of these fluctuations and eliminates an explanation in terms of classical 
hybridization effects.
Finally, our experimental findings are in quantitative agreement with a microscopic
theory based on the self-consistent harmonic approximation (SCHA), embedded within
a fully-fledged microscopic description of our circuit using microwave simulation tools.

\begin{figure*}[htb]
\begin{center}
\includegraphics[width = 0.8\textwidth]{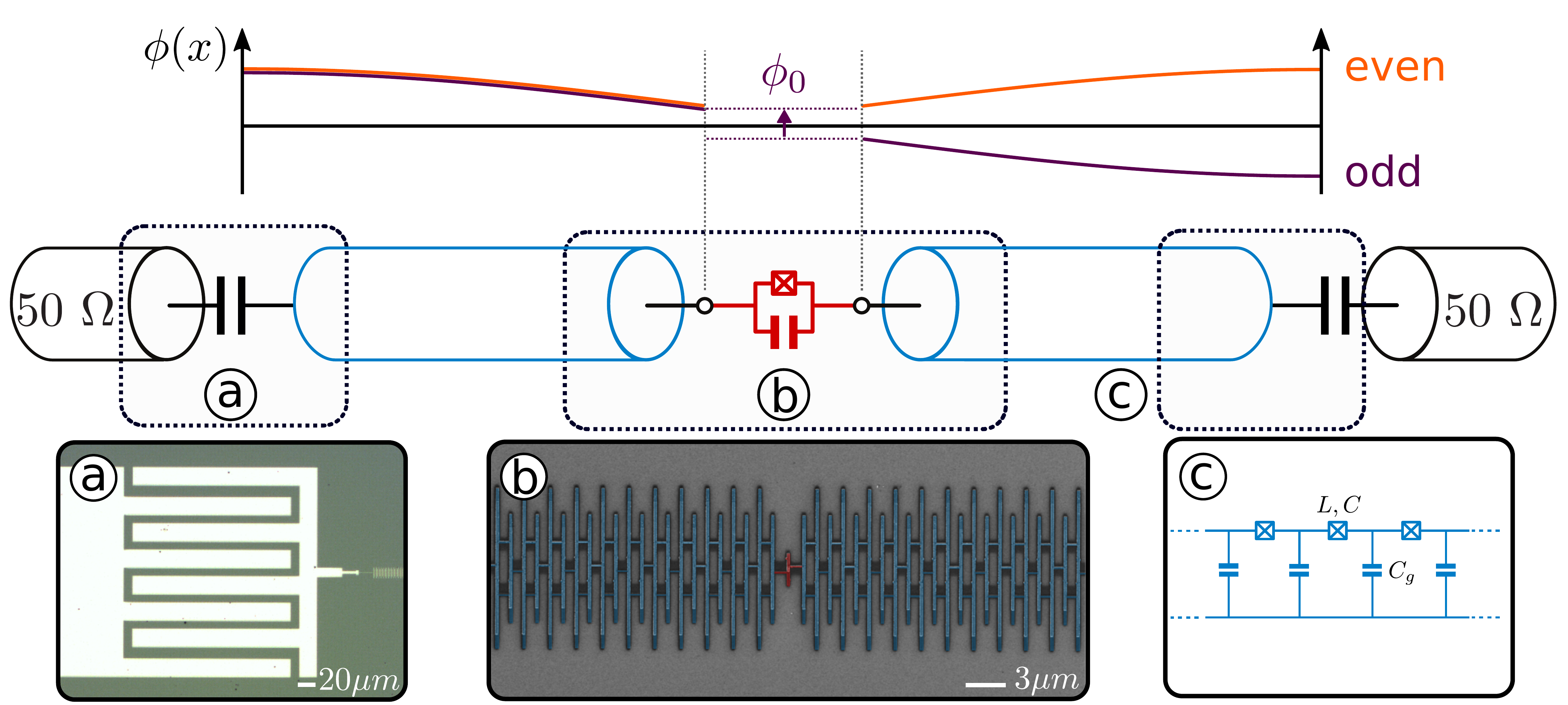}
\caption{\textbf{SQUID chains coupled to a small Josephson junction (weak link).} 
The upper part represents the spatial phase distribution of the two first standing waves
or resonant modes of the total system (Josephson junction + chains). And odd (even)
mode -- which couples (does not couple) to the junction -- is represented in
purple (orange). The lower part is a schematic of the system. The SQUID chains, depicted
as blue transmission lines, are capacitively coupled to the input and output 50
Ohm coaxial cables and galvanically coupled to the small Josephson junction (in
red). \textbf{a} Optical picture of the input and output capacitive
couplings. \textbf{b} SEM picture of a few of the SQUIDs (1500 in total for
each chain) that are coupled to the small Josephson junction (in red). 
\textbf{c} Equivalence between the transmission line effective picture and the 
SQUID chain characterized by three microscopic parameters $L$ and $C$ the inductance 
and capacitance per SQUID respectively and $C_g$ the ground capacitance. 
}
\label{fig1}
\end{center}
\end{figure*}

\noindent{\large \textbf{Results} \par}
\noindent\textbf{Background}\\
The many-body regime of a single non-linear junction coupled to a high
impedance environment has remained largely unexplored experimentally, since obtaining
$Z_\text{env}\sim R_\text{Q}$ at gigahertz frequencies is very challenging. One option is
to use on-chip resistors~\cite{Kuzmin:1991ev}. However, this is may lead to
unwanted Joule heating~\cite{Huard2007}.
Therefore, we rather pursue a solution that relies on superconducting (lossless) high
inductance materials such as Josephson junction
arrays~\cite{Manucharyan2012,Masluk:2012jo,Bell:2012co}, noting that disordered
superconductors~\cite{Maleeva2018} are also promising. In Josephson junction arrays,
$Z_\text{env} = \sqrt{L/C}$ can reach $R_\text{Q}$ given the large inductance of these
materials, while maintaining good quality factors in the device.

Early experiments have embedded ultra-small Josephson junctions between highly
resistive leads, demonstrating the incoherent tunneling of Cooper pairs~\cite{Kuzmin:1991ev}
in the framework of the $P(E)$ theory~\cite{Ingold:1992dt}. In this case however, no
supercurrent flows through the junction and no quantum coherent effects were observed.
Later, the phase/charge duality in the regime $Z_\text{J}$,
$Z_\text{env}$ > $R_\text{Q}$ was explored using SQUID arrays as the
environment~\cite{Corlevi:2006vu,Ergul:2013fr,Weissl:2015bj}.
Experimental results were explained by fluctuations due to the finite
temperature of the electromagnetic environment and the effect of zero-point
fluctuations could not be investigated. Moreover, these two series of
experiments relied on DC measurements. This has the disadvantage that
non-equilibrium effects need to be taken into account when results are interpreted,
while the system is not directly probed at the finite frequencies
 -- around $\omega_\text{J}$ -- that are of greatest interest.

It has since become possible to obtain a frequency-resolved picture of the
environment of quantum systems such as Josephson junctions, thanks to the advent
of circuit QED~\cite{Wallraff:2004rV}. Here, microwave techniques allow a more
accurate examination of the effects of zero point fluctuations on Josephson
junctions~\cite{Hoi:2015fh}, and observations of perturbative spectral shifts (below 1\%) attributed to ZPF were
reported~\cite{Fragner:2008ct,Silveri:2019cxa,Wen:2019us}. Several bottom-up experiments
explored nonperturbative effects of light-matter interaction at
ultra-strong coupling between a qubit and a single-mode resonator (for a review
see \cite{FornReview,Kockum:2019ky}). An effect similar to the Lamb shift -- a reduction of the
effective Josephson energy -- was also reported recently for a DC-biased Josephson
junction coupled to a single mode high impedance resonator~\cite{Rolland:2019fo}.
Moving towards many-body territory, a non-perturbative renormalization of
the frequency of a flux qubit was demonstrated~\cite{FornDiaz:2016bo,Magazzu:2018is}.
However, in this experiment, fluctuations were mainly thermal, and in addition, the
environment cutoff frequency could not be clearly measured. The resulting unknown
parameters prevented a quantitative modeling of the experiments.
Indeed, as pointed by various
authors~\cite{GarciaRipoll:2015ba, Malekakhlagh:2017ca, Gely:2017cwa,Parra_Rodriguez_2018}, it is necessary to account for all the microscopic details of the circuit to get rid of unphysical divergences in multi-mode models. Furthermore, a thorough modeling of such circuits is mandatory to discriminate the trivial effects of normal mode splitting (spectral shifts observed when two classical harmonic oscillators hybridize) from the dynamical ones associated to true vacuum fluctuations. With the exception of~\cite{Gely:2018cw},  this important issue has received surprisingly little attention in the circuit QED context.

\noindent\textbf{Presentation of the experiment}\\
Our system builds on recent advances in the fabrication and control of 
large-scale Josephson arrays~\cite{PuertasMartinez:2019gk,Kuzmin:2019cg}. 
It consists of a small Josephson junction of characteristic impedance on the 
order of $R_\text{Q}$ ($E_\text{J}/E_\text{c} \lesssim 1$), which is embedded 
in the middle of two SQUID chains, each consisting of 1500 unit cells (figure~\ref{fig1}), 
forming high characteristic impedance transmission lines. We measure the 
characteristics of this environment  precisely : its high frequency cut-off -- or plasma frequency --
$\omega_\text{plasma} \simeq $ \SI{17}{\giga\hertz} 
 and its wave impedance $Z_\text{chain}=\sqrt{L/C_\text{g}} \simeq$ \SI{1.8}{\kilo\ohm} 
(see table~\ref{table} and
Supplementary Note 10). The SQUID parameters were carefully chosen to maintain a
negligible phase slip rate ($E_\text{J}/E_\text{c} \lesssim 500$), ensuring that
these chains can be described as a linear environment. They are capacitively
coupled to the measurement setup to suppress DC noise which could affect the
small junction (figure \ref{fig1}.\textbf{c}). In order to vary the degree of non-linearity
and hence the strength of the ZPF, we measured three samples with
different small junction sizes, connected to nominally identical chains (see
table~\ref{table}).

\begin{figure*}[htb!]
\begin{center}
\includegraphics[width = 0.8\textwidth]{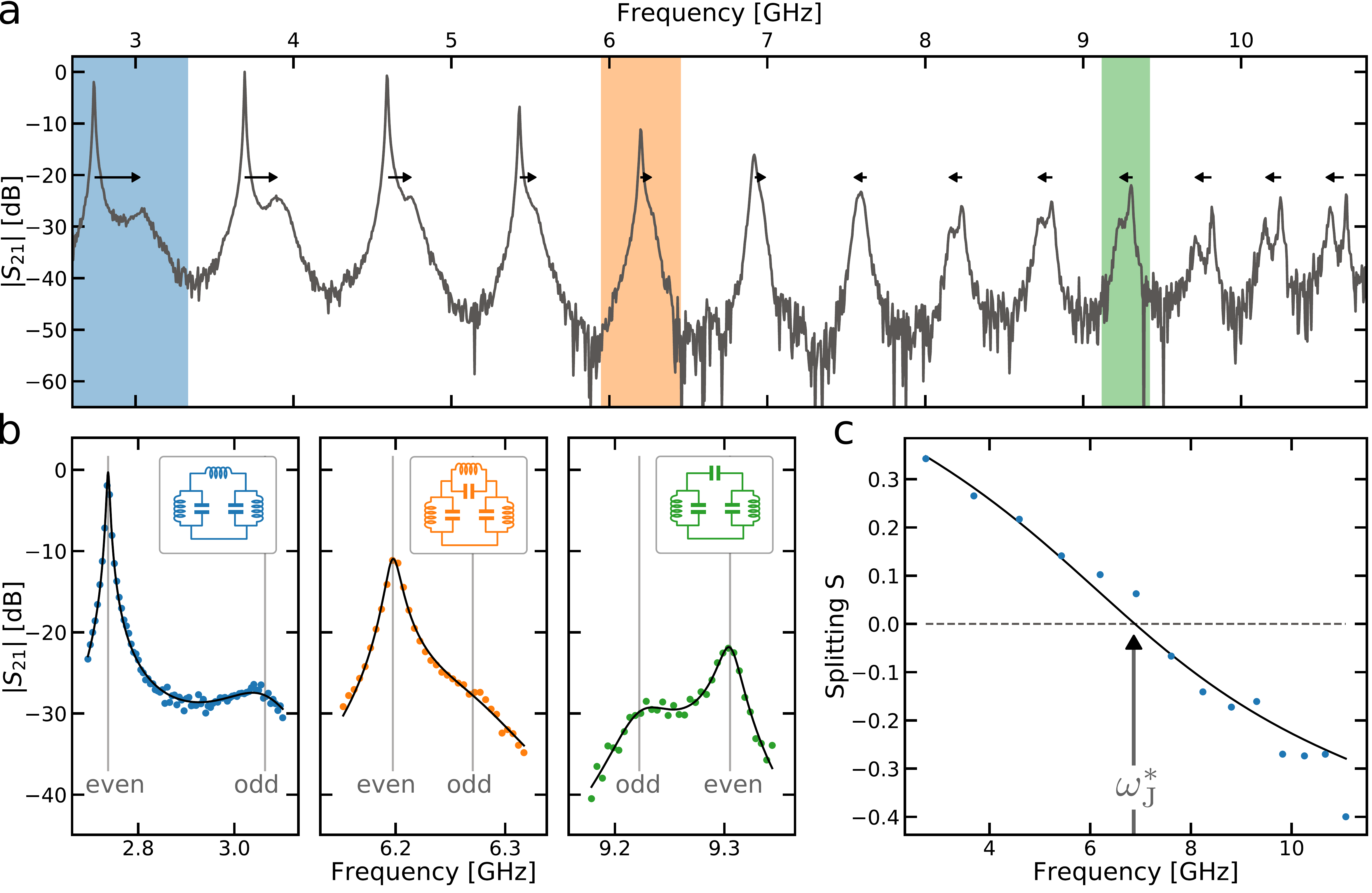}
\caption{\textbf{Inferring the renormalized resonant frequency
$\omega_\text{J}^*$ of the small Josephson junction.} 
\textbf{a} Amplitude of the microwave transmission $|S_{21}|$ versus frequency
(sample A, 24mK). The even-odd modes frequency splitting $S$ changes sign
precisely at $\omega_\text{J}^*$. Arrows are guides to the eye of the splitting sign.
\textbf{b} Fit of the double peaks for three cases: well below the resonant frequency 
of the small Josephson junction (blue) its inductive part dominates, close to $\omega_\text{J}^*$ 
(orange) the impedance of the junction is large so that the two modes are almost
decoupled, and well above $\omega_\text{J}^*$ (green) the capacitive part of the junction 
dominates.
\textbf{c} Experimental normalized frequency splittings $S$ obtained from the previous fits (dots)
and theoretical prediction (full line). The resonance frequency
$\omega_\text{J}^*$ of the small Josephson junction corresponds to the vanishing
value of the normalized splitting $S$.
}
\label{fig2}
\end{center}
\end{figure*}

The broadband microwave transmission of the full system shows a series of
resonances (see figure~\ref{fig2}.\textbf{b}). A broadening of the modes 
in the array is expected since
the SQUID chains are capacitively coupled to the \SI{50}{\ohm} measurement lines,
hence forming very long microwave resonators. The transmission of
the system is measured using very low microwave power, down to the single photon regime. This prevents any 
power-induced broadening or frequency shift of these resonances (see Supplementary Note 9). 
A closer look at figure~\ref{fig2}.\textbf{b} reveals that 
resonances come in pairs. This is expected given the symmetry of the sample:
our system can be decomposed into two subsystems (See Supplementary Note 2). One
is made of even modes, which are decoupled from the small Josephson junction,
while the other is composed of odd modes, with impedance $Z_\text{env} = 2
Z_\text{chain}$, ultra-strongly coupled to the small Josephson
junction~\cite{PuertasMartinez:2019gk,Kuzmin:2019cg}. A more surprising
observation is that the odd modes are much more damped than the even ones. 
We interpret this as resulting from the non-linearity that odd modes inherit from the small 
Josephson junction. This is experimental evidence of the strong 
back-action of the small Josephson junction on the many modes of the chain forming 
its linear environment.

\begin{table}[!h]
\caption{\label{param_samples}\textbf{Parameters of three samples.}
The bare Josephson energy $E_\text{J,bare}^\text{AB}$ is inferred using the Ambegaokar-Baratoff 
law. $E_\text{J}^*$ is the measured value of the renormalized Josephson energy.
As a consistency check, the bare value $E_\text{J,bare}^\text{th}$ is also extracted from 
the fit of $E_\text{J}^*$ using the SCHA. $C_\text{sh}$ is the capacitance shunting the small Josephson junction
(see Supplementary Note 9). $C$, $C_\text{g}$ and $L$ are obtained from the
dispersion relation of the chain (see Supplementary Note 10).}

\begin{tabular}{|c|ccc|}
\hline
Sample                     & A            & B             & C \\
\hline

Small junction &&&\\
\hline

Area [$\mu m^2$]                  & 315x195        & 370x190         & 440x185\\
$C_\text{J}$ [fF]                 &2.7 $\pm$ 0.3 & 3.2 $\pm$0.3  & 3.7 $\pm$0.4 \\
$C_\text{sh}$ [fF]                &3.0 $\pm$ 0.5 & 2.4 $\pm$ 0.4 & 5.1 $\pm$ 1.0 \\
$E_\text{J}^* $[GHz]              &1.8 $\pm$ 0.1 & 3.1 $\pm$ 0.2 & 5.7 $\pm$ 0.3\\
$E_\text{J,bare}^\text{AB}$ [GHz] &3.7 $\pm$ 0.2 & 5.8 $\pm$ 0.3 & 6.8 $\pm$ 0.5\\
$E_\text{J,bare}^\text{th}$ [GHz] & 3.7          & 5.5            & 8.2\\
\hline 
Non-linearity $E_\text{J,bare}/E_\text{c}$                         &0.27           & 0.40            & 0.93\\
Renormalization $E_\text{J}^*/E_\text{J,bare}$                       &0.49           & 0.56            & 0.70\\
\hline 

Chain &&&\\
\hline

$C $ [fF]                 & 144            & 144             & 144\\
$C_{\text{g}}$ [fF]                 & 0.189          & 0.192           & 0.181\\
$L$ [nH]                   & 0.66           & 0.60            & 0.61\\
\hline 
$E_\text{J}/E_\text{c}$                  & 460           &  506            & 498\\
\hline
\end{tabular}

\label{table}
\end{table} 

\noindent\textbf{Line shapes}\\
The line shape of a given even-odd pair of resonances can be obtained
by associating with it two effective LC oscillator~\cite{pozar2009microwave}
connected via the small Josephson 
junction (see insets in figure~\ref{fig2}b).
In the regime of interest (See Supplementary Note 5 and 6) this junction can be treated
as a ZPF-dependent inductance $L_\text{J}^*$ in parallel with a capacitance $C_{\parallelsum}$,
with resonance frequency~$\omega_\text{J}^* =
1/\sqrt{L_\text{J}^*C_{\parallelsum}}$.
The odd and even modes mentioned earlier are
characterized by respective frequencies $\omega_\text{even} =
1/\sqrt{LC}$ and $\omega_\text{odd} = 1/\sqrt{L_{\Sigma}C_{\Sigma}}$, with
$1/L_{\Sigma} = 1/2L + 1/L_\text{J}^*$ and $C_\Sigma = C/2 + C_{\parallelsum}$. Then
for modes at frequencies such that $\omega_\text{odd},\omega_\text{even}\ll
\omega_\text{J}^*$, the capacitance of the small junction can be neglected
($C_{\parallelsum}\sim 0$) leading to $\omega_\text{odd}>\omega_\text{even}$. In the
opposite case ($\omega_\text{odd},\omega_\text{even}\gg \omega_\text{J}^*$) the
inductance can be neglected, giving $\omega_\text{odd}<\omega_\text{even}$. The
most interesting regime is when the system is probed close to
$\omega_\text{J}^*$. In that case, the impedance of the small junction 
diverges and consequently the two effective oscillators are uncoupled leading
to $\omega_\text{odd} = \omega_\text{even}$. In the Supplementary
Note 7, we confirm that a fully microscopic model of the whole circuit also predicts 
that the frequency splitting between even and odd modes changes sign at the renormalized 
frequency of the junction $\omega_\text{J}^*$. 
The frequencies of each even-odd pair of modes is extracted by fitting the peaks Fig.~\ref{fig2} to
line shapes of an input-output formalism based on the simple model just described (see 
Supplementary Figure 3 and Methods).

\noindent\textbf{Renormalized Josephson energy $E_\text{J}^*$}\\
The effective resonance frequency of the junction, $\omega_\text{J}^*$,
depends on its environment due to the interplay of strong anharmonicity and
many-body ZPF, and can be inferred by tracking the evolution of the normalized
frequency splitting
$S=(\omega_\text{odd,k}-\omega_\text{even,k})/(\omega_\text{even,k+1}-\omega_\text{even,k})$,
between even (uncoupled) and odd (coupled) modes,
where $\text{k}=0\ldots M$ refers to mode number. 
As shown in Ref. ~\cite{PuertasMartinez:2019gk}, in a long chain, this quantity
equals the phase shift difference between even and odd modes. It vanishes when
the left and right halves of the device decouple, so that even and odd modes
become degenerate.  Figure~\ref{fig2}.\textbf{c} shows the experimentally
obtained $S$ for one of our samples, from which we extract $\omega_\text{J}^*$. 
As we show in the Supplementary Note 4 and 5, the ZPF-dependent effective inductance of the weak link is related to a renormalized 
Josephson energy $E_\text{J}^*=(\hbar/2e)^2/L_\text{J}^*$ as 
\begin{equation}
	\omega_\text{J}^* = \sqrt{2E_\text{J}^* E_\text{c}},  
	\label{eq_wjstart}
\end{equation}
where $E_\text{c} = (2e)^2/(2(C_\text{J}+C_\text{sh}))$, with $C_\text{J}$ the intrinsic capacitance of the junction and
$C_\text{sh}$ a shunting capacitance due to the surrounding circuitry.  Note that we  define $E_\text{J}^*$ in terms of
$L_{\text{J}}^*$, and not in terms of the DC critical current as is done in for instance~\cite{Hekking:1997fs}.
We use Eq.\,(\ref{eq_wjstart}) to infer $E_\text{J}^*$ experimentally. 
$C_\text{J}$ is given by the junction size measured from an SEM picture. The way
$C_\text{sh}$ is extracted is explained in the Supplementary Note 9. Values for sample A, B and C are reported
in table \ref{param_samples}.
To see the effect of vacuum fluctuations, we compare $E_\text{J}^*$ to the bare
Josephson energy of the weak link, which was obtained as follows.
We fabricated many nominally identical Josephson junctions on the same
chip and measured their room temperature resistances. The expected bare
Josephson energy of the small Josephson junction $E_\text{J,bare}^\text{AB}$
(see table \ref{param_samples}) was then inferred using the Ambegaokar-Baratoff law.
We observe a systematic shift between this bare energy and the renormalized one we inferred from
$|S_{21}|$ measurements, a shift that is more pronounced for sample A that
shows a high non-linearity. This points towards a large renormalization induced by
the strong zero-point phase fluctuations of the hybridyzed junction-chain modes, as expected
since the small junction is impedance-matched to the chains. 

\begin{figure}[ht]
\begin{center}
\includegraphics[scale = 0.35]{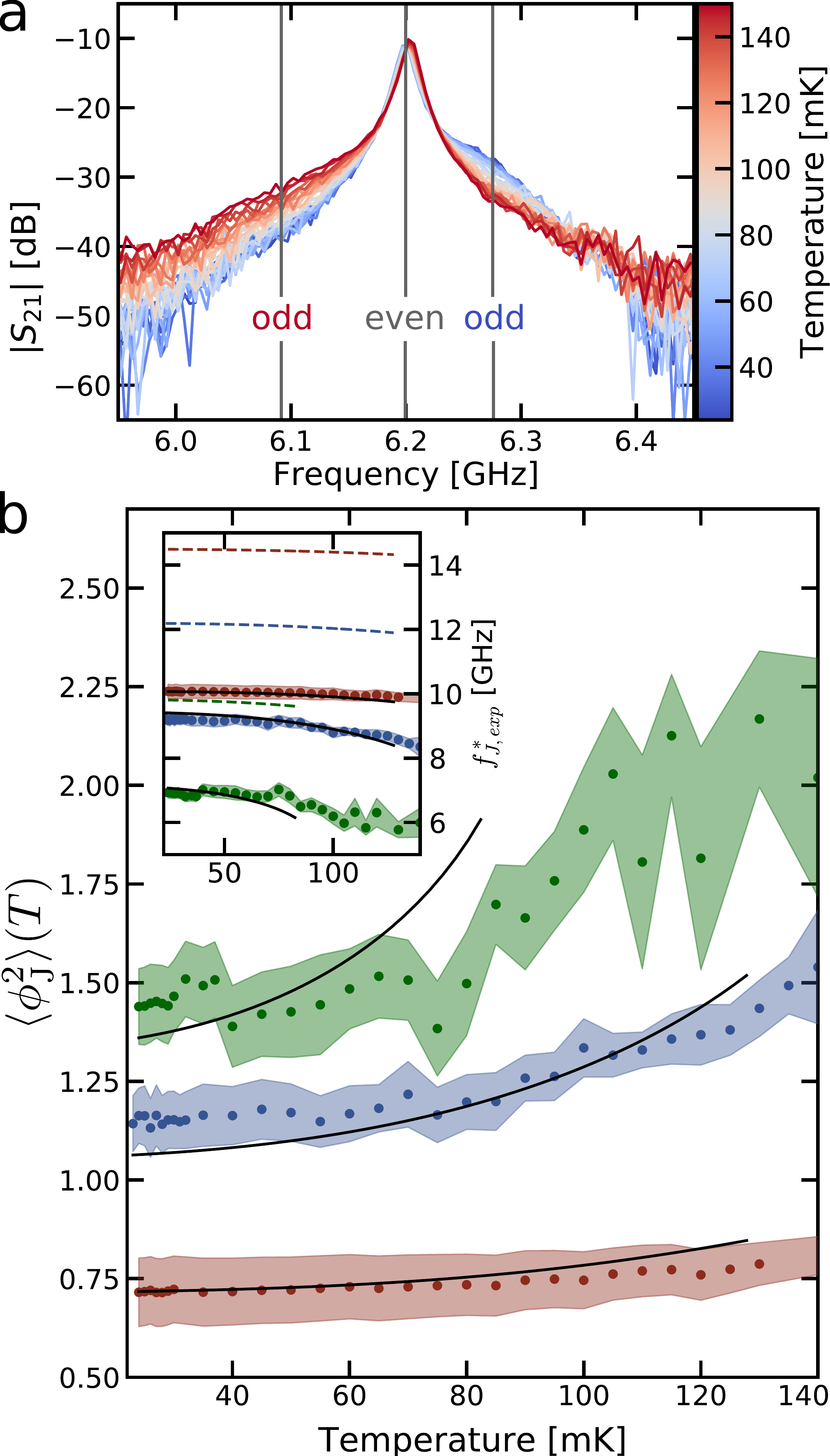}
\caption{\textbf{Temperature-induced renormalization.}
\textbf{a} Zoom on a even-odd pair of transmission peaks for sample A at temperature ranging from 23 to 150 mK. The even mode (gray) does not move while the odd mode (blue is at 25mK, red at 130mK)  shift down in frequency when
warming up, showing a downward renormalization of the junction frequency $\omega_\text{J}^*$. 
\textbf{b} ZPF of the small junction $\expval{\phi_\text{J}^2}$ as a function of the
temperature for three samples (A,B and C ranging from dark to light blue), extracted from Eq.~(\ref{self-consistent}). 
ZPF are stronger in sample A, which is associated to a smaller ratio
$E_\text{J,bare}/E_\text{c}$ (large non-linearity). The measured quantum to classical 
crossover is in good agreement with theory (full lines). The inset displays the corresponding renormalised junction frequency $f_\text{J}^*=\omega_\text{J}^*/2\pi$
of the three samples. The full lines are the SCHA predictions while the dashed lines represent what would be the temperature evolution of these frequencies if ZPF were omitted from $\expval{\phi_\text{J}^2}$, using the same values of $E_\text{J,bare}$.
}
\label{fig3}
\end{center}
\end{figure}

We now show that this renormalization is quantitatively captured by a
microscopic model based on the self consistent harmonic approximation 
(SCHA). Its success in accounting for nonlinearities introduced by Josephson junctions 
is well-established~\cite{Zaikin:1986,Panyukov:1988,Kampf:1987,Chakravarty:1988}.
More recently it was employed in detailed microscopic models in the field of circuit-QED~\cite{PuertasMartinez:2019gk,Joyez:2013fl}.
The idea behind the 
SCHA is that the strong phase fluctuations allowed by the environment 
average the non-linear potential of the small Josephson junction, lowering 
its effective Josephson energy from the bare value $E_\text{J}$ to the 
renormalized one $E_\text{J}^*$. 
This is valid, provided the phase $\phi_\text{J}$, though strongly fluctuating, is still sufficiently localized.
In this regard we note the following. Though large, the effective environmental impedence $2 Z_\text{chain}\simeq$\SI{3.8}{\kilo\ohm}   
seen by the weak link, is still less than $R_\text{Q}$. 
Under this condition, the environment is known to produce spontaneous symmetry breaking of the 2$\pi$ periodicity in the phase difference $\phi_\text{J}$ across the weak link, 
\cite{Schmid:1983vi,Lett:wq,Schon:1990kj}. 
It is therefore reasonable to approximate the system's full wave function with a Gaussian that is fairly well localized in the $\phi_\text{J}$ direction, which is the essence of the SCHA.
At zero
temperature, the interplay of many-body ZPF and non-linearity can be described is approximated  
by replacing the cosine Josephson potential by an effective quadratic term 
$E_\text{J}^*\phi_\text{J}^2/2$, where the renormalized Josephson energy 
$E_\text{J}^*$ is given by the self-consistent equation:
\begin{equation}
E_\text{J}^* = E_\text{J,bare}\, e^{-\expval{\phi_\text{J}^2(E_\text{J}^*)}/2}.
\label{self-consistent}
\end{equation}

Here, the total phase fluctuation across the junction
$\expval{\phi_\text{J}^2}$ is given by: 
\begin{equation}
	\expval{\phi_\text{J}^2} = \sum_{k \in \text{odd}}{\phi_k^2},
\end{equation}
where $\phi_k^2$ is the contribution to the small junction ZPF
coming from odd mode $k$. Importantly, in the strong ZPF regime, the expectation
value must be taken with respect to the modified vacuum of the hybridized modes,
which means that the normal modes of the systems has to be updated during the
numerical iteration of Eq.~(\ref{self-consistent}).
This is in contrast to familiar examples of ZPF induced phenomena, such as the
Lamb shift in hydrogen, where the perturbative nature of the effect allows one to
calculate fluctuations with respect to the bare vacuum of the environment.
We independently extracted the parameters of the whole circuit (junction+chains), and then used Eq.\,(\ref{self-consistent}) to
determine the theoretical bare Josephson energy required to 
find back the measured renormalized $E_\text{J}^*$ (see next section for more details). The agreement between experimentally and theoretically estimated
$E_\text{J,bare}$ (see table \ref{param_samples}) provides strong evidence that our system displays large ZPF, which leads to a
renormalization of up to $50 \%$ of the Josephson energy of the small
junction (or equivalently $30\%$ of its resonant frequency $\omega_\text{J}$).
Moreover, as expected, this renormalization increases when the ratio
$E_\text{J,bare}/E_\text{c}$ decreases, or equivalently when the non-linearity of
the small Josephson junction increases.\\ 

\noindent\textbf{Quantum versus thermal fluctuations}\\
As $\omega_\text{J}^*$ is renormalized by phase fluctuations across the weak link, 
one expects a crossover from quantum to thermally driven fluctuations as temperature 
increases. Extending the SCHA to non-zero temperatures (see Supplementary
Note 4 and 5), we find that the fluctuations of mode $k$ contain a Bose factor
contribution:
\begin{equation}
\phi_k^2(T) = \phi_k^2 
\left[1+\frac{2}{\mathrm{exp}(\hbar\omega_k/k_\text{B}T)-1}\right]
\label{thermalZPF}
 \end{equation}
with $\omega_k$ the frequency of mode $k$, and $\phi_k^2$ its
zero temperature ZPF. Therefore, at low temperature, fluctuations saturate
to a finite ZPF value (a hallmark of quantum uncertainty), while at high temperature they 
increase linearly with temperature~(\ref{thermalZPF}).
According to Eq. (\ref{self-consistent}), $\omega_\text{J}^*$ should decrease when the system is heated up. Consequently, odd modes' frequencies are shifted to lower values when temperature increases, while the even modes stay put. This striking experimental signature of non-linearity can clearly be seen in \ref{fig3}.\textbf{a}. This constitutes smoking gun evidence of the back-action of the Josephson junction on its environment: the shift of $\omega_\text{J}^*$ to smaller values at increasing temperatures indicates that fluctuations are thermally enhanced.

The recipe to extract $E_\text{J,bare}^\text{th}$ is the following: $E_\text{J}^*(T)$ is obtained from $S_{21}$ measurements at different temperatures. Since all the other parameters ($L$, $C$, $C_\text{g}$, $C_\text{J}$ and $C_\text{sh}$) are known, we can fit $E_\text{J}^*(T)$ using Eqs. \eqref{self-consistent} and \eqref{thermalZPF}, taking $E_\text{J,bare}^\text{th}$ as the (only) fitting parameter. Then, $E_\text{J,bare}^\text{th}$ being determined, we can compute the phase fluctuations across the small Josephson junction using Eq. (\ref{eq_wjstart}) and (\ref{self-consistent}): 
\begin{equation}
	\expval{\phi_\text{J}^2}(T) = 4\log(\frac{2E_\text{J,bare}^\text{th}E_\text{c}}{\omega_J^*(T)}).
\end{equation}
We checked that at the lowest temperature of our cryostat, the phase fluctuations experienced by the small Josephson junction are fully in the quantum regime, by measuring $|S_\text{21}|$ from 25mK to 130mK. Results are shown in figure~\ref{fig3}.\textbf{b}. We observe that the quantum to classical crossover appears at decreasing temperatures from sample C to A. This is because $\omega_\text{J}^*$ decreases from sample C to A. Therefore the junction is coupled to modes with lower and lower frequencies, which are thermally occupied at lower temperatures. The inset in figure~\ref{fig3}.\textbf{b} shows the corresponding fit of $\omega_\text{J}^*$ for the three samples. The dashed lines represent $\omega_\text{J}^*$ obtained using the value of $E_\text{J,bare}$ extracted from the previous fit but including only thermal renormalization of $\omega_\text{J}^*$ i.e. disregarding ZPF. 
Consequently, $\expval{\phi_\text{J}^2}$ is given by:
\begin{equation}
\phi_k^2(T) = \phi_k^2 
\left[\frac{2}{\mathrm{exp}(\hbar\omega_k/k_\text{B}T)-1}\right]
\label{thermalZPF_2}
\end{equation}
The discrepancy between the dashed lines and the fit clearly shows that
the fluctuations have mainly quantum origin. At increasing temperatures, thermal
fluctuations add up to the quantum ZPF, and cause a rise in
$\big<\phi_\text{J}^2\big>$, witnessed both in the experimental extraction and
the predictions from SCHA, see figure~\ref{fig3}\textbf{b}. 
It is likely that the extracted $\big<\phi_\text{J}^2\big>$ for sample A is systematically underestimated
due to sizeable errors in the SCHA that rapidly set in after $\big<\phi_\text{J}^2\big>\gtrsim 1$, leading to a mismatch with the 
theory at high temperatures.

\begin{figure}[htb]
\begin{center}
\includegraphics[scale=0.35]{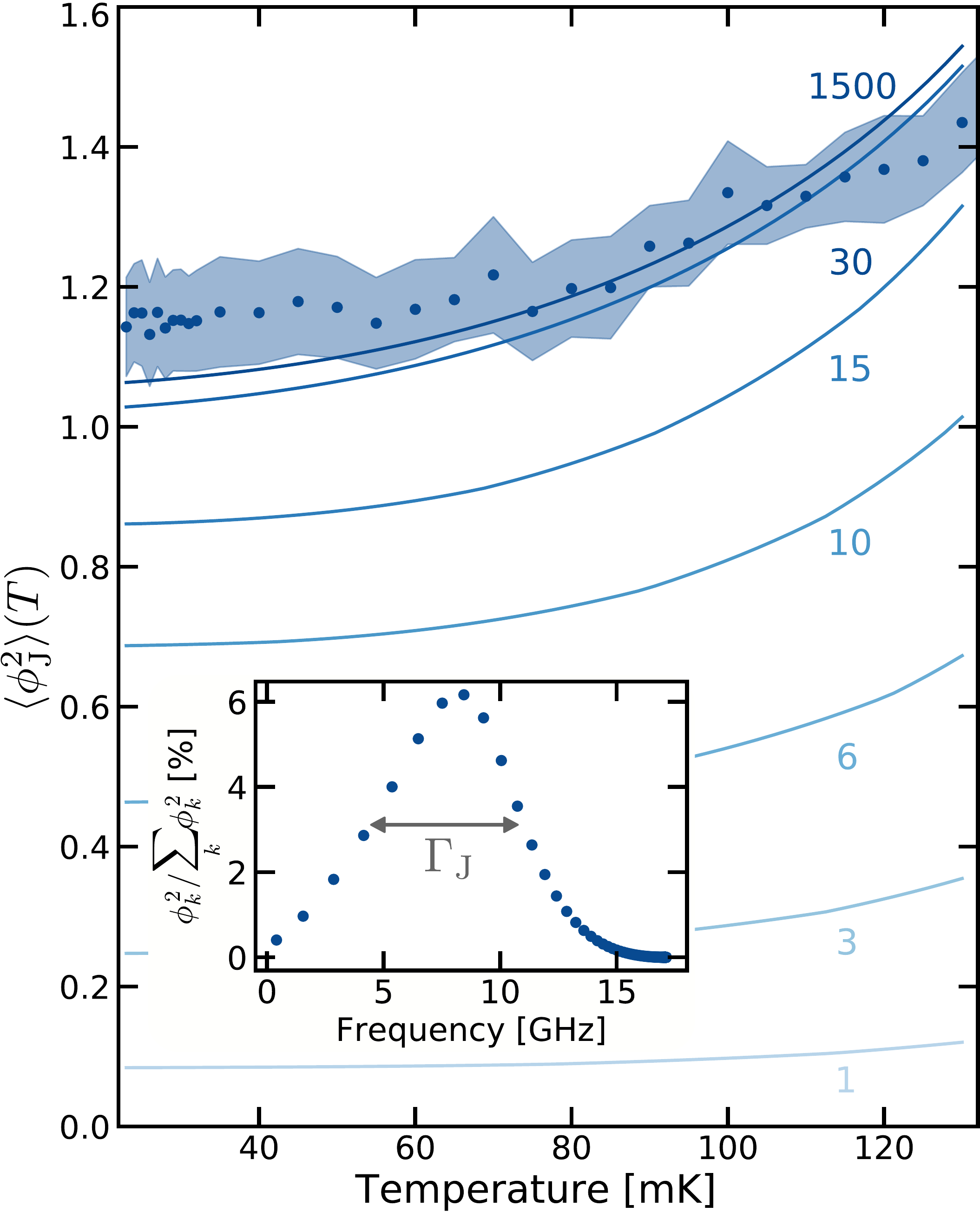}
\caption{\textbf{Many-body nature of the ZPF.}
 Total phase fluctuations across the small Josephson junction
in sample B, taking into account in our model (full lines)
different numbers of modes of the environment, ranging from one (light blue) to the total number (dark blue). 
The inset shows the relative contribution of the different modes to the total
fluctuations, with $\Gamma_\text{J}$ being the FWHM of this quantity. 
}
\label{fig4}
\end{center}
\end{figure}

\noindent\textbf{Many-body nature of the ZPF}\\
In order to confirm the many-body character of this renormalization, we
can estimate how many modes are affecting the small junction simultaneously.
The ZPF are quantitatively determined by how the full vacuum of the whole circuit is dressed
by the coupling through the weak-link. Within the SCHA,
the number of modes contributing a finite amount of $\phi_k^2$ provides
a measure of the number of interacting particles in the system.
In figure \ref{fig4}. we compare the experimentally extracted
$\expval{\phi_\text{J}^2(T)}$ to various calculated values. In each calculation,
the full system was truncated to a finite number of modes in a window around $\omega_\text{J}^*$.
If the window is too narrow, important contributions to the ZPF are neglected, and
$\expval{\phi_\text{J}^2(T)}$ is underestimated.   
The comparison unambiguously shows that, in sample B, around 30 modes
contribute to the total phase fluctuations. In circuit QED language, the full
width at half maximum (FWHM) of the environmental ZPF $\phi_\text{k}^2
(\omega_\text{k})$ -- labeled $\Gamma_\text{J}$ -- is about 7GHz for our samples
(see inset of figure \ref{fig4}.). Therefore, our device operates in a regime where
$\Gamma_\text{J}/\omega_\text{J}^* \sim 1$ due to the impedance matching to the transmission line. Moreover, our device is strongly non-linear. 
Consequently, it is not possible to treat perturbatively the non-linearity
as is usually done in the field for the Transmon qubit or other weakly non-linear circuits~\cite{Koch:2007gz,Nigg:2012ek,Bourassa2012,Weissl:2015do}
(a detailed analysis is given in Supplementary Note 11).\\

This work provides a direct observation of several quantum many-body effects
driven by zero point fluctuations in an open quantum system. This was achieved by developing a spectroscopic setup where the high-impedance environment 
of a single non-linear Josephson junction was monitored mode by mode, and compared to a detailed microscopic model.
A strong quantum renormalization (up to 50\%) of the Josephson energy of the
single junction was demonstrated, analogous to a non-perturbative Lamb shift.
In addition, the back-action of the small Josephson junction causes
non-linear broadening and strong temperature dependence of the environmental
modes, providing the most striking signature of the many-body effects that 
take place in our circuit.
The measured temperature dependence of the phase fluctuation across the
Josephson junction indicates that our device remains quantum coherent at
cryogenic temperatures. As many as 30 modes are involved in the renormalization 
of the small junction. Our superconducting circuit thus behaves as a fully fledged 
quantum many body simulator, paving the way for the further observation of various
many-body non-linear effects in circuit-QED
~\cite{GarciaRipoll:2008il,LeHur:2012wq,Goldstein:2013kq,Peropadre:2013iaa,Snyman:2015dm,Gheeraert:2017gq,Gheeraert:2018bv}.\\

\normalsize
\noindent{\bf Correspondence}\\
\footnotesize
\noindent
Correspondence and requests for materials should be addressed to Nicolas 
Roch~(email: nicolas.roch@neel.cnrs.fr).\\

\normalsize
\noindent{\bf Methods}\\
\footnotesize

\noindent{\bf Full Model}\\
\noindent
The Hamiltonian of the full system can be decomposed into odd and even
parts -- containing respectively the modes coupled and not coupled to the
junction (see Supplementary Note 2). The odd Hamiltonian reads:
\begin{eqnarray}
	\hat{H} &=& \hat{H}_0 + \left(1 - E_\text{J,bare}\cos\hat\phi_\text{J}\right), 
\label{hamil_odd}\\
	\hat{H}_0 &=& \frac{\left(2e\right)^2}{2}\sum_{i,j = 0}^{N}{\hat{n}_{i}
\big[ \hat{C}\big]^{-1}_{i,j}\hat{n}_{j}}+ \frac{E_\text{J,S}}{4}
\sum_{i = 1}^{N-1}{\left(\hat{\phi}_{i} - \hat{\phi}_{i+1}\right)^2},
\end{eqnarray}
with $\hat{n}_{0}\equiv \hat{n}_{J}$
and $\hat\phi_\text{J}$ referring to the charge
the phase drop across the small junction
while $\hat{n}_{i}$ and $\hat{\phi}_{i}$, $i \in [1..N]$ refer to the charge and phase operators on chain site
$i \in [1..N]$. Charge and phase operators obey the commutation rules
$\big[\hat{\phi}_{k},\hat{n}_{p}\big] = i\delta_{k,p}$.  The microscopic parameters are 
$E_\text{J,S}$, the Josephson energy of the SQUIDs, $E_\text{J,bare}$ the bare 
Josephson energy of the small junction, and the capacitance matrix:
 $$\hat{C}=\frac{1}{2}
\begin{bmatrix}
C_I    & -C         & 0         &0       &0          &\dots      & 0  \\
-C     &   2C+C_g   &     -C    &  0     & 0         & \dots     & 0  \\
0      &   -C       & 2C+C_g    & -C     & 0         & \dots     & 0  \\
\vdots & \vdots     & \ddots    & \ddots & \ddots    & \dots     & 0  \\ 
0      &   0        &   0       &   -C   & 2C+C_g    &  -C       & 0  \\ 
0      &   0        &   0       &    0   &  -C       &  2C+C_g   &-C  \\ 
0      &   0        &   0       &    0   &   0       &   -C      & C_0\\
\end{bmatrix}
$$
with : 
\begin{flalign}
	C_I &= 2(C_\text{J} + C_\text{sh}) + C + C_\text{g},\\
	C_0 &= C_\text{c} + C_\text{c,I} + C.
\end{flalign}
\noindent{\bf Self Consistent Harmonic Approximation (SCHA)}\\
\noindent
Because of the cosine term in Eq.~(\ref{hamil_odd}), we are dealing with an interacting
many-body problem that cannot be solved analytically. To study the best variational harmonic approximation we use the SCHA:
\begin{flalign}
\hat{H} &= \hat{H}_0 + \frac{E^*_\text{J}}{2}\hat\phi_\text{J}^2 
+ \left(1 - E_\text{J,bare}\cos\hat\phi_\text{J}\right) -
\frac{E^*_\text{J}}{2}\hat\phi_\text{J}^2 \\
	&= \hat{H}_\text{t} + \left(1 - E_\text{J,bare}\cos\hat\phi_\text{J}\right) 
- \frac{E^*_\text{J}}{2}\hat\phi_\text{J}^2,
\end{flalign}
with $\hat{H}_\text{t}$ the trial harmonic Hamiltonian that will approximate $\hat{H}$,
optimized with respect to the renormalized Josephson energy $E^*_\text{J}$. The variational
principle gives: 
\begin{equation}
	\frac{\partial}{\partial E^*_\text{J}} \expval{\hat{H}}{\Psi_\text{t}} = 0
\label{var_pri},
\end{equation}
with $\ket{\Psi_\text{t}}$ the many-body ground state of $\hat{H}_\text{t}$. Because of
the harmonic character of $\hat{H}_\text{t}$, we have:
\begin{equation}
\expval{\cos\hat\phi_\text{J}}{\Psi_\text{t}} = e^{-\expval{\hat\phi_\text{J}^2}{\Psi_\text{t}}/2}
\label{harm_exp}.
\end{equation}
Inserting (\ref{harm_exp}) into (\ref{var_pri}) we end up with the self consistent equation: 
\begin{equation}
E_\text{J}^* = E_\text{J,bare}e^{-\expval{\phi_\text{J}^2(E_\text{J}^*)}_\text{t}/2}.
\end{equation}
The physical interpretation is the following: when ZPF are
negligible, $\big<\hat\phi_\text{J}^2\big> \simeq 0$ and $E_\text{J}^* = E_\text{J,bare}$, 
so in its ground state the junction behaves as an harmonic oscillator of frequency 
$\omega_\text{J,bare} = \sqrt{2E_\text{J,bare}E_\text{c}}$. 
For weak non-linearity, fluctuations increase but remain such that 
$\big<\hat\phi_\text{J}^2\big> \ll 1$, resulting in
$E_\text{J}^* \simeq E_\text{J,bare}(1 - \big<\phi_\text{J}^2\big>/2)$, so that the
junction behaves as a weakly anharmonic oscillator with fundamental frequency
$\omega_\text{J,bare}(1 - \big<\hat\phi_\text{J}^2\big>/4)$. For an isolated
junction $\big<\hat\phi_\text{J}^2\big> = \sqrt{E_\text{c}/2E_\text{J,bare}}$,
and the frequency becomes $\omega_\text{J,bare} - E_\text{c}/4$, a well known
result for the Transmon qubit~\cite{Koch:2007gz}. For larger fluctuations, the
principle remains the same but no analytical formula can be derived, so that one
should solve the self consistent equation numerically. A more detailed derivation -- 
including thermal fluctuations -- is presented in Supplementary Note 4 and 5.
 
\noindent{\bf Frequency splitting S between odd and even modes}\\
\noindent
The splitting $S$ is linked to the phase shift difference $\theta$ between even and odd modes in the thermodynamic
limit~\cite{PuertasMartinez:2019gk}: 
\begin{equation}
	S = \frac{\theta}{\pi}.
\end{equation}
The analytical formula of the phase shift difference-- derived in the Supplementary 6 and 7 -- reads : 
\begin{equation}
\theta = 2\, {\rm arccot}(X)+{\rm arctan}\left[\frac{1-\lambda}{1+\lambda}X\right]
\end{equation}
with 
\begin{flalign}
X &= \sqrt{\left(\frac{4C}{C_\text{g}}+1\right)\left(\left(\frac{\omega_\text{p}}{\omega}\right)^2-1\right)}, \\
\lambda &= \frac{1 - \omega^2CL}{1 + 2L/L_\text{J}^* - \omega^2C_\text{I} L}, \label{phase_shift_csh}
\end{flalign}
$\omega_\text{p}=1/\sqrt{L(C+C_\text{g}/4)}$ being the plasma frequency of the chain and
$L_\text{J}^*=\hbar^2/(2e)^2E_\text{J}^*$ the effective inductance of the small junction.

\noindent{\bf Fitting formula for the peaks}\\
\noindent
Input-output theory is used to fit the parameters associated to the double
resonances observed in the transmission spectrum. These are mapped to two
coupled harmonic modes $\alpha$ and $\beta$ with mutual coupling rate $g$,
external coupling $\kappa_\text{ext}$ and internal loss $\kappa_\text{in}$, with 
Hamiltonian:
\begin{equation}
\hat{H} = \hbar\omega_\text{r}(\hat{a_\text{L}}^\dagger\hat{a_\text{L}} 
+ {a_\text{R}}^\dagger\hat{a_\text{R}}) +
g(\hat{a_\text{L}}+\hat{a_\text{L}}^\dagger)(\hat{a_\text{R}}+\hat{a_\text{R}}^\dagger).
\end{equation}
Here $\hat{a}_{\text{in}_{\text{L}}}$,$\hat{a}_{\text{out}_{\text{L}}}$ are the left input and output signals and
$\hat{a}_{\text{out}_{\text{R}}}$ is the right output signal. Thus the input-output
relations are :
\begin{flalign}
\hat{a}_{\text{in}_{\text{L}}} + \hat{a}_{\text{out}_{\text{L}}} &= \sqrt{\kappa_\text{ext}}\hat{a_\text{L}}
\label{eq_in1},\\ 
\hat{a}_{\text{out}_{\text{R}}} &= \sqrt{\kappa_\text{ext}}\hat{a_\text{R}}. \label{eq_in2}
\end{flalign} 
The equations of motion are:
\begin{flalign}
-i(\omega - i\omega_\text{r})\hat{a_\text{L}} + \frac{\kappa_\text{ext}}{2}\hat{a_\text{L}}&= -ig\hat{a_\text{R}}
- \sqrt{\kappa_\text{ext}} \hat{a}_{\text{in}_{\text{L}}}, \label{eq_motion1}\\
-i(\omega - i\omega_\text{r})\hat{a_\text{R}} + \frac{\kappa_\text{ext}}{2}\hat{a_\text{R}}&= -ig\hat{a_\text{L}}. \label{eq_motion2}
\end{flalign}
The complex transmission is defined as 
$S_{21} = \hat{a}_{\text{out}_{\text{R}}}/\hat{a}_{\text{in}_{\text{L}}}$, and can be calculated using Eqs. (\ref{eq_in1}-\ref{eq_motion2}).
We define the even 
$\omega_\text{e} = \omega_\text{r}+g$
and odd $\omega_\text{o} = \omega_\text{r}-g$ frequencies, 
and add phenomenologically losses in the odd modes 
$\kappa_\text{o} = \kappa_\text{in} + \kappa_\text{add}$ (we keep $\kappa_\text{e} 
= \kappa_\text{in}$), so that: 
\begin{equation}
S_{21} = \frac{i\kappa_\text{ext}(\omega_\text{o} - \omega_\text{e})}{\left(\kappa_\text{ext}+\kappa_\text{o} +
-2i(\omega-\omega_\text{o})\right)\left(\kappa_\text{ext}+\kappa_\text{e}
-2i(\omega-\omega_\text{e})\right)}.
\end{equation}
For some of the odd modes, we found a signature of inhomogeneous
broadening, that we modeled by a convolution of their frequency with a gate
function defined as $\Pi_{\delta\omega}(\omega) = 1/\delta\omega$ if $\omega \in
[\omega - \delta\omega/2, \omega + \delta\omega/2]$.  
Understanding microscopically this additional broadening, possibly due to offset
charges, is beyond the scope of the description using the SCHA, and will require
additional theoretical developments.

\normalsize

\noindent{\bf Data Availability}\\
\footnotesize
\noindent
The data that support the findings of this study are available from the corresponding
author upon reasonable request.\\

\normalsize
\noindent{\bf Acknowledgements}\\
\footnotesize
\noindent
The authors would like to thank F. Balestro, L. Del Rey, D. Dufeu, E. Eyraud, J.
Jarreau, T. Meunier and W. Wernsdorfer, for early support with the experimental
setup. Very fruitful discussions with K. R. Amin, P. Forn-Diaz, J.-J. Garcia-Ripoll, 
D. B. Haviland, M. Houzet, P. Joyez, V. E. Manucharyan, F. Portier and H. E. Tureci 
are acknowledged. The sample was fabricated in the Nanofab clean room. This research 
was supported by the ANR under contracts CLOUD (project number ANR-16-CE24-0005), 
GEARED (project number ANR-14-CE26-0018), by the National Research Foundation of 
South Africa (Grant No. 90657), and by the PICS contract FERMICATS. J.P.M. acknowledges 
support from the Laboratoire d\textquoteright excellence LANEF in Grenoble
(ANR-10-LABX-51-01). R.D. and S.L. acknowledge support from the CFM
foundation and the 'Investisements d'avenir'  (ANR-15-IDEX-02) programs of the French 
National Research Agency. K.B. and J.D. acknowledge the European Union's Horizon 2020 research and innovation programme under the Marie Sklodowska-Curie grant agreement No 754303.\\

\normalsize
\noindent{\bf Competing interests}\\
\footnotesize
\noindent
The authors declare no competing financial or non-financial interests.\\

\normalsize
\noindent{\bf Author contributions}\\
\footnotesize
\noindent
S.L., J.P.M., S.F. and N.R. designed the experiment. S.L. fabricated the 
device. S.L. performed the experiment and analysed the data with 
help from S.F., N.R. and I.S., while S.F. and I.S. provided the theoretical 
support. S.L., J.P.M., K.B., R.D., J.D., F.F., V.M., L.P., O.B., C.N., W.H.G., S.F., I.S. and N.R. participated in setting up the experimental platform, and took part in writing the paper.\\


%

\newpage
\setcounter{figure}{0}
\setcounter{table}{0}
\setcounter{equation}{0}

\onecolumngrid

\global\long\def\theequation{S\arabic{equation}}
\global\long\def\thefigure{S\arabic{figure}}
\renewcommand{\thetable}{S\arabic{table}}
\renewcommand{\arraystretch}{0.6}

\normalsize

\vspace{1.0cm}
\begin{center}
{\bf \large Supplementary information for
"Observation of quantum many-body effects due to zero point fluctuations in superconducting circuits''}
\end{center}

\subsection{Supplementary Note 1: Experimental setup}

The measurement setup is displayed in Supplementary Figure~\ref{meas_setup}. The samples are put
in a dilution refrigerator with a 25mK base temperature. $|S_{21}|$ is
measured using a Vector Network Analyzer (VNA). An additional microwave
source was used for two-tone measurements, while a global magnetic field was
applied via an external superconducting coil. Both the coil and the sample were
held inside a mu-metal magnetic shield coated on the inside with a light
absorber made out of epoxy loaded with silicon and carbon powder. IR filters are
0.40mm thick stainless steel coaxial cables. The bandwidth of the measurement setup
goes from 2.5 GHz to 12 GHz.

\begin{figure}[h]
\includegraphics[width=0.75\textwidth]{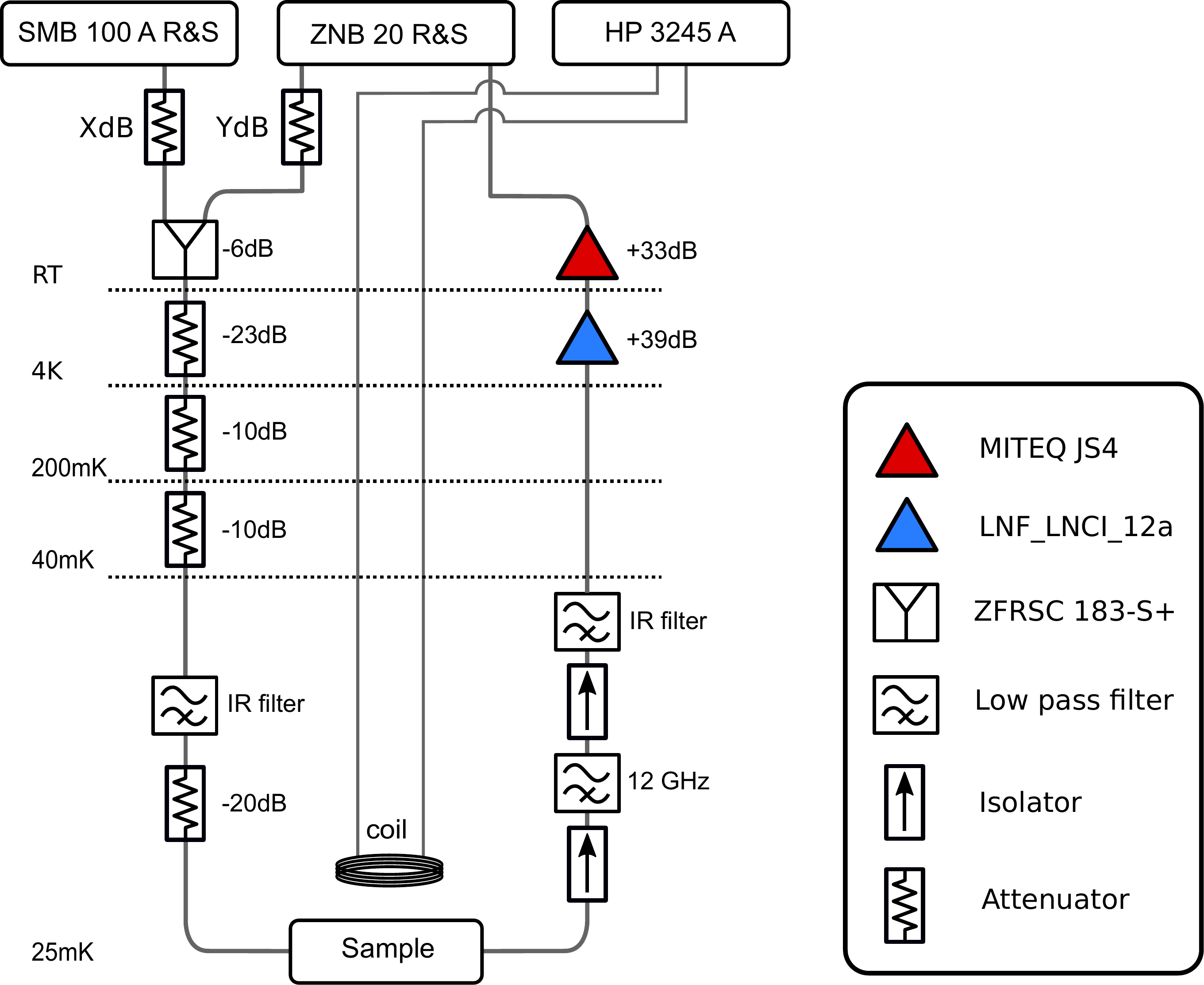}
\caption{{\bf Measurement setup.}}
\label{meas_setup}
\end{figure}

\subsection{Supplementary Note 2: Odd and Even modes}

Our device consists of two long Josephson chains of $N+1$ sites tailored in the 
linear regime (with Josephson energy $(\hbar/2e)^2/L$ much larger than the capacitive energy)
interconnected via a smaller Josephson junction or weak-link (operating in the regime of
small Josephson energy $E_{J,\text{bare}}$). Linearizing the tunneling term within each chain,
but keeping the non-linear coupling between them, the Hamiltonian of the system
reads:
\begin{equation}
\hat{H} = \frac{\left(2e\right)^2}{2}\sum_{i,j = 0}^{N}\sum_{\sigma,\sigma'\in L,R}
{\hat{n}_{i\sigma}[C]^{-1}_{i,\sigma,j,\sigma'}\hat{n}_{j,\sigma'}} 
+ \frac{1}{2}\frac{\hbar^2}{(2e)^2L}\sum_{i = 1}^{N-1}\sum_{\sigma \in L,R}
{\left(\hat{\phi}_{i,\sigma} - \hat{\phi}_{i+1,\sigma}\right)^2} 
- E_{J,\text{bare}}\cos\left(\hat{\phi}_{0,L} - \hat{\phi}_{0,R}\right),
\end{equation}
with $\hat{n}_{i,\sigma}$ and $\hat{\phi}_{i,\sigma}$ the charge and phase
operators on site $i \in [1..N]$ and in chain $\sigma=L,R$. These operators are canonically 
conjugate and obey at the quantum mechanical level the commutation rules
$\big[\hat{\phi}_{i,\sigma},\hat{n}_{j,\sigma'}\big] =
i\delta_{i,j}\delta_{\sigma,\sigma'}$. The capacitance matrices can be read off
the equivalent circuit in Supplementary Figure~\ref{Fig:circuit}, and are decomposed into an
intra-chain part $[C_0] = [C]_{LL} = [C]_{RR}$ and an interchain part
intra-chain part $[C_1] = [C]_{LR} = [C]_{RL}$, which read explicitely:

$$
[C_0] = 
\begin{bmatrix}
C_\text{I}    &   -C              &      0            &  0     & 0                & \dots            & 0  \\
-C            &   2C+C_\text{g}   &     -C            &  0     & 0                & \dots            & 0  \\
0             &   -C              &  2C+C_\text{g}    & -C     & 0                & \dots            & 0  \\
\vdots        & \vdots            & \ddots            & \ddots & \ddots           & \dots            & 0  \\ 
0             &   0               &   0               &   -C   & 2C+C_\text{g}    &  -C              & 0  \\ 
0             &   0               &   0               &    0   &  -C              &  2C+C_\text{g}   &-C  \\ 
0             &   0               &   0               &    0   &   0              &   -C             & C_\text{O}\\
\end{bmatrix}
,
$$
and $[C_1]_{i,j} = -\delta_{0,i}\delta_{j,0}
(C_\text{J}+C_\text{sh})$ with  $(i,j) \in [0,N]^2$. 
The total capacitance at the weak-link end of the chain amounts 
to $C_\text{I} = C_\text{J}+C_\text{sh}+C+C_\text{g}$, while the capacitance at
the connecting output port is $C_\text{O} = C_\text{c}+C_\text{c,I}+C$.
\begin{figure}[!h]
\includegraphics[width=\textwidth]{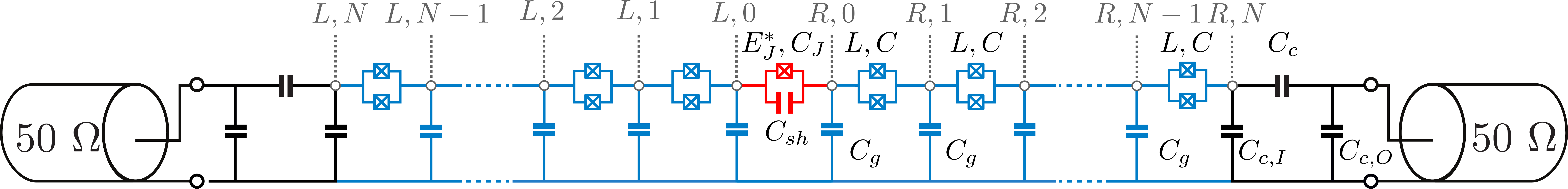}
\caption{ {\bf Electrical circuit of the device.} The capacitance network is
indicated for the output ports (in black), the two chains (in blue) and the weak
link (in red).}
\label{Fig:circuit}
\end{figure}

Due to the symmetry of our device, it is useful to define respectively even and odd modes:
\begin{flalign}
\hat{n}_{j,\pm} &= \frac{1}{2}\left(\hat{n}_{j,R} \pm \hat{n}_{j,L} \right),\\
\hat{\phi}_{j,\pm} &= \left(\hat{\phi_{j,R}} \pm \hat{\phi_{j,L}} \right).
\end{flalign}
In this basis, the Hamiltonian decomposes in two uncoupled subsystems:
$\hat{H} = \hat{H}_+ + \hat{H}_-$, where:
\begin{flalign}
\hat{H}_+ &=  \frac{\left(2e\right)^2}{2}\sum_{i,j = 0}^{N}{\hat{n}_{i,+}
\left[\frac{C_0+C_1}{2}\right]^{-1}_{i,j}\hat{n}_{j,+}} 
+ \frac{1}{4}\frac{\hbar^2}{(2e)^2L}\sum_{i = 1}^{N-1}{\left(\hat{\phi}_{i,+} 
- \hat{\phi}_{i+1,+}\right)^2},\\ 
\hat{H}_- &=  \frac{\left(2e\right)^2}{2}\sum_{i,j = 0}^{N}
{\hat{n}_{i,-}\left[\frac{C_0-C_1}{2}\right]^{-1}_{i,j}\hat{n}_{j,-}}+
\frac{1}{4}\frac{\hbar^2}{(2e)^2L}\sum_{i = 1}^{N-1}{\left(\hat{\phi}_{i,-} -
\hat{\phi}_{i+1,-}\right)^2} + E_\text{J} \left(1 - \cos\hat{\phi}_{0,-}\right).
\end{flalign}
$\hat{H}_+$ reduces to the Hamiltonian of a linear chain, while $\hat{H}_-$ 
takes the form of a boundary Sine-Gordon-like model.

\subsection{Supplementary Note 3: Fitting the transmission resonances}

The transmission spectrum consists of pairs of peaks, that are fitted according
to the model described in the Methods section of the main text. Close to a pair
of even/odd resonances, the transmission is given by the formula:
\begin{equation}
S_{21} = \frac{i\kappa_\text{ext}(\omega_\text{o} - \omega_\text{e})}{\left(\kappa_\text{ext}+\kappa_\text{o} +
-2i(\omega-\omega_\text{o})\right)\left(\kappa_\text{ext}+\kappa_\text{e}
-2i(\omega-\omega_\text{e})\right)},
\label{eq:FitS12}
\end{equation}
with $\omega_\text{o}$ and $\omega_\text{e}$ the even/odd resonance frequencies, $\kappa_\text{e}$
and $\kappa_\text{o}$ their respective intrinsic damping rate, and $\kappa_\text{ext}$ the
broadening due to the 50 $\Omega$ output ports. A large selection of fitted
spectra (for all three samples and various temperatures) is shown in
Supplementary Figure~\ref{Fig:Fitting}.

\begin{figure}[H]
\begin{center}
\includegraphics[width=\textwidth]{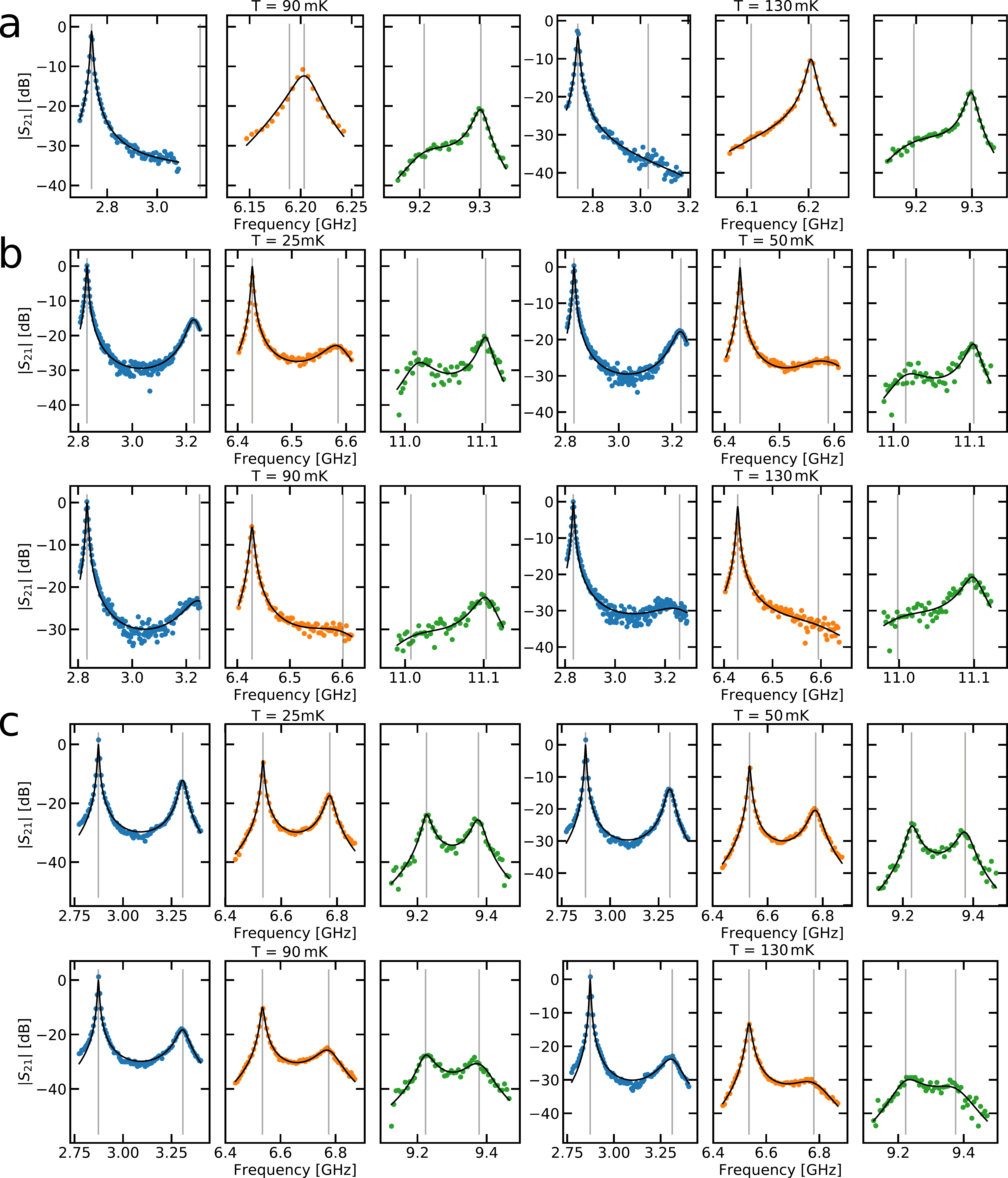}
\caption{ 
{\bf Fitting of various peak pairs}. Panel \textbf{a} is for sample 300,
panel \textbf{b} is for sample 375 and panel \textbf{c} is for sample 450. 
Various temperature choices are indicated, and for each case, three frequencies
ranges are indicated (in blue, orange and green respectively).
Vertical grey lines are the positions of the resonance pairs found by the
regression (black lines) using formula~(\ref{eq:FitS12}).}
\label{Fig:Fitting}
\end{center}
\end{figure}

\subsection{Supplementary Note 4: The self consistent harmonic approximation}

The hamiltonian $\hat{H}_-$ describes a quantum many-body problem that cannot be solved analytically, and
we therefore develop here an approximate yet microscopic approach to the
problem. From now on, we will discard the - index in all fields, and replace 
$\hat{\phi}_{0,-}$ by $\hat{\phi}_\text{J}$. 
The self consistent harmonic approximation (SCHA) is used to find 
the approximate ground state at thermal equilibrium~\cite{SJoyez:2013fl,SPuertasMartinez:2019gk}. This method consist of finding 
the best harmonic Hamiltonian $\hat{H}_\text{t}$ which satisfies the Gibbs-Bogoliubov inequality
$F \leq F_\text{t} + \langle \hat{H} - \hat{H}_\text{t} \rangle_\text{t}$,
where: 
\begin{flalign}
F_\text{t} &= -k_\text{B}T\ln Z_\text{t}, \\ 
Z_\text{t} &= \tr(e^{-\hat{H}_\text{t}/k_\text{B}T}), \\
\langle \hat{H} - \hat{H}_\text{t} \rangle_\text{t}  &= \tr((\hat{H} - \hat{H}_\text{t})\hat{\rho}_\text{t}), \\
\hat{\rho}_\text{t}  &= \frac{1}{Z_\text{t}}e^{-\hat{H}_\text{t}/k_\text{B}T}.
\end{flalign}
The trial Hamiltonian $\hat{H}_\text{t}$ is defined by replacing in $\hat{H}$ the non-linear
tunneling term $-E_\text{J}\cos\hat{\phi}_\text{J}$ by a renormalized potential
$E_\text{J}^*\hat{\phi}_\text{J}^2/2$. The physical reason is that the zero point fluctuations
of the small junction explore a large part of the Josephson potential, which
amounts in first approximation to lower its effective Josephson energy from the
bare value $E_\text{J}$ to a renormalized value $E_\text{J}^*$. Explicitely, the trial
Hamiltonian reads:
\begin{equation}
\hat{H}_\text{t}  =\frac{\left(2e\right)^2}{2}\sum_{i,j = 0}^{N}
{\hat{n}_{i}[C]^{-1}_{i,j}\hat{n}_{j}} 
+ \frac{1}{2}\frac{\hbar^2}{(2e)^2}\sum_{i,j = 0}^{N}{\hat{\phi}_{i}[L^{-1}]_{i,j}\hat{\phi}_{j}},
\end{equation}
with the capacitance matrix: 
$$
[C] = \frac{1}{2}
\begin{bmatrix}
C_{\Sigma}    &   -C              &  0                 &  0        & 0                & \dots            & 0  \\
-C                 &   2C+C_\text{g}   &     -C             &  0        & 0                & \dots            & 0  \\
0                  &   -C              & 2C+C_\text{g}      & -C        & 0                & \dots            & 0  \\
\vdots             & \vdots            & \ddots             & \ddots    & \ddots           & \dots            & 0  \\ 
0                  &   0               &   0                &   -C      & 2C+C_\text{g}    &  -C              & 0  \\ 
0                  &   0               &   0                &    0      &  -C              &  2C+C_\text{g}   &-C  \\ 
0                  &   0               &   0                &    0      &   0              &   -C             & C_\text{O}\\
\end{bmatrix},
$$
where $C_\Sigma=C_\text{I}+C_\text{J}+C_\text{sh}=2(C_\text{J}+C_\text{sh})+C+C_\text{g}$, and inductance matrix: 
$$
[L^{-1}] = \frac{1}{2}
\begin{bmatrix}
2/L^*+1/L   & -1/L       &  0        &  0     & 0         & \dots     & 0    \\
-1/L        & 2/L        & -1/L      &  0     & 0         & \dots     & 0    \\
0           & -1/L       & 2/L       & -1/L   & 0         & \dots     & 0    \\
\vdots & \vdots     & \ddots    & \ddots & \ddots         & \dots     & 0    \\ 
0           &   0        &   0       & -1/L   & 2/L       & -1/L      & 0    \\ 
0           &   0        &   0       &    0   & -1/L      & 2/L       &-1/L  \\ 
0           &   0        &   0       &    0   &   0       &   -1/L    & 1/L  \\
\end{bmatrix}.
$$
Here $L^*=(\hbar/2e)^2/E_\text{J}^*$ is an effective inductance
associated with the weak link.
 
Let us define by {$E_k = \hbar\omega_k$} the eigenvalues of $\hat{H}_\text{t}$
and $\hat{a}_k^\dagger$ the corresponding creation operators associated to its 
normal modes. As $\hat{H}_\text{t}$ is harmonic, one can write:
\begin{flalign}
\label{eq:Htharmo}
\hat{H}_\text{t}  &=\sum_{k = 0}^{N+1}{\hbar\omega_k \hat{a}_k^\dagger \hat{a}_k},\\
\hat{\phi}_\text{J} &= \sum_{k = 0}^{N+1}{\phi_k(\hat{a}_k^\dagger +  \hat{a}_k)} \label{eq_phi_diag}.
\end{flalign}
The renormalized Josephson energy $E_\text{J}^*$ is obtained by minimizing the
variational free energy:
\begin{equation}
\frac{d}{dE_\text{J}^*}(F_\text{t} + \langle \hat{H} - \hat{H}_\text{t} \rangle_\text{t}) = 0.
\label{eq_minimize}
\end{equation}
The first term is evaluated as follows:
\begin{flalign}
\dfrac{dF_\text{t}}{dE_\text{J}^*} &= - \frac{k_\text{B}T}{Z_\text{t}}\dfrac{dZ_\text{t}}{dE_\text{J}^*}
= - \frac{k_\text{B}T}{Z_\text{t}}\sum_{k}{\dfrac{d}{dE_\text{J}^*}\left(e^{-E_k/k_\text{B}T}\right)}
=\frac{1}{Z_\text{t}}\sum_{k}\bra{k}\dfrac{d\hat{H}_\text{t}}{dE_\text{J}^*}\ket{k}e^{-E_k/k_\text{B}T}\\
&=  \frac{1}{Z_\text{t}}\sum_{k}{\bra{k}\frac{\hat{\phi}_\text{J}^2}{2}\ket{k}e^{-E_k/k_\text{B}T}}
=   \frac{\langle \hat{\phi}_\text{J}^2\rangle_\text{t}}{2} \label{gb_t1}
\end{flalign}
where we used the fact that $\bra{k}\hat{H}_\text{t} \dfrac{d}{dE_\text{J}^*}\ket{k}=0$, which follows because $\ket{k}$ is a normalized eigenstate of $\hat{H}_\text{t}$ and $\dfrac{d}{dE_\text{J}^*}\ket{k}$ 
is orthogonal to $\ket{k}$.
The second term in the variational free energy is 
\begin{equation}
\frac{d}{dE_\text{J}^*}\langle \hat{H} - \hat{H}_\text{t} \rangle_\text{t} = - \frac{E_{\text{J},\text{bare}}}{2}\dfrac{d}{dE_\text{J}^*}\langle
e^{i\hat{\phi}_\text{J}} + e^{-i\hat{\phi}_\text{J}} \rangle_\text{t} - \frac{\langle
\hat{\phi}_\text{J}^2\rangle_\text{t}}{2} -
\frac{E_\text{J}^*}{2}\dfrac{d}{dE_\text{J}^*}\langle\hat{\phi}_\text{J}^2\rangle_\text{t} \label{gb_t2}.
\end{equation}
Inserting Eq.~(\ref{gb_t1}) and Eq.~(\ref{gb_t2}) in Eq.~(\ref{eq_minimize}),
one finds the following condition on $E_\text{J}^*$ : 
\begin{equation}
E_\text{J}^* = -E_\text{J,bare}\frac{\dfrac{d}{dE_\text{J}^*}(\langle e^{i\hat{\phi}_\text{J}} +
e^{-i\hat{\phi}_\text{J}} \rangle_\text{t})}{\dfrac{d}{dE_\text{J}^*}\langle\hat{\phi}_\text{J}^2\rangle_\text{t}}.
\label{eq:SCHAint}
\end{equation}

\subsection{Supplementary Note 5: Microscopic model}

Let us now compute $\langle e^{i\hat{\phi}_\text{J}}\rangle_\text{t}$ using
Eq.~(\ref{eq_phi_diag}) and the Baker-Campbell-Hausdorff formula : 
\begin{flalign}
\langle e^{i\hat{\phi}_\text{J}}\rangle_\text{t} &= \langle \exp(i\sum_{k = 0}^{M}{\phi_k(\hat{a}_k^\dagger +  \hat{a}_k)})\rangle_\text{t}
= \langle \exp(i\sum_{k = 0}^{M}{\phi_k\hat{a}_k^\dagger})  \exp(i\sum_{k = 0}^{M}{\phi_k\hat{a}_k})               \rangle_\text{t}\exp(-\frac{1}{2}\sum_{k = 0}^{M}\phi_k^2) \\
&=\langle \sum_{n\geq 0}\sum_{m\geq 0}{\frac{i^ni^m}{n!m!}\left(\sum_{k = 0}^{M}
{\phi_k\hat{a}_k}\right)^n\left(\sum_{k = 0}^{M}
{\phi_k\hat{a}_k}\right)^m}\rangle_t\exp(-\frac{1}{2}\sum_{k = 0}^{M}\phi_k^2).
\end{flalign}
The terms where $n=m$ are the only one different from 0 : 
\begin{flalign}
\langle e^{i\hat{\phi}_\text{J}}\rangle_\text{t} &= \sum_{m\geq 0}{\frac{\left(-1\right)^n}{n!^2}\sum_{k_1...k_n}\phi_{k_1}...\phi_{k_n}\sum_{k_1'...k_n'}\phi_{k_1'}...\phi_{k_n'}\langle a_{k_1}^\dagger...a_{k_n}^\dagger a_{k_1'}...a_{k_n'}\rangle_t}  \label{wicks_before} \\
&=\sum_{n\geq 0}\frac{\left(-1\right)^n}{n!^2}n!\sum_{k_1...k_n}{\phi_{k_1}^2...\phi_{k_n}^2\langle a_{k_1}^\dagger a_{k_1} \rangle_t...\langle a_{k_n}^\dagger a_{k_n} \rangle_t}\exp(-\frac{1}{2}\sum_{k = 0}^{N}\phi_k^2)\label{wicks_after} \\
&= \sum_{n\geq 0}{\frac{1}{n!}\left(-\sum_{k = 0}^{N}\left(n_k\phi_k^2 \right)\right)^n}\exp(-\frac{1}{2}\sum_{k = 0}^{N}\phi_k^2)
=\exp(-\sum_{k = 0}^{N}(n_k+\frac{1}{2})\phi_k^2)
=\exp(-\langle \hat{\phi}_\text{J}^2 \rangle_\text{t} / 2).
\end{flalign}
Wick's theorem has been used between Eq.~(\ref{wicks_before}) and
Eq.~(\ref{wicks_after}), and $n_k=1/[\exp(\hbar\omega_k/k_\text{B}T)-1]$ is the Bose factor. 
One verifies easily that $\langle e^{-i\hat{\phi}_\text{J}}\rangle_\text{t} = \langle e^{i\hat{\phi}_\text{J}}\rangle_\text{t}$. 
We can finally simplify the term appearing in Eq.~(\ref{eq:SCHAint}):
\begin{equation}
\dfrac{d}{dE_\text{J}^*}\langle e^{i\hat{\phi}_\text{J}} + e^{-i\hat{\phi}_\text{J}} \rangle_\text{t} = - e^{-\langle\hat{\phi}_\text{J}^2\rangle_\text{t}}	\dfrac{d}{dE_\text{J}^*}(\langle \hat{\phi}_J^2 \rangle_\text{t})
\end{equation}
so that $E_\text{J}^*$ obeys the simple self-consistency relation:
\begin{equation}
\boxed{
E_\text{J}^* =E_\text{J,bare}e^{-\langle\hat{\phi}_\text{J}^2\rangle_\text{t}/2} 
= E_\text{J,bare}\exp[-\sum_{k = 0}^{N}\phi_k^2(n_k+\tfrac{1}{2})]}\label{eq:boxedSCHA}
\end{equation}

We finally present the procedure to compute the normal mode expansion
coefficients $\phi_k$, as obtained from the trial Hamiltonian
$\hat{H}_\text{t}$. The original charge and phase variables can be decomposed
formally onto the normal modes:
\begin{flalign}
\hat{n}_p   &= -\frac{i}{2\pi}\sqrt{\frac{R_Q}{2}}\sum_{k = 1}^{N}[G]_{p,k}(\hat{a}_k-\hat{a}_k^\dagger) \label{Q_mode},\\ 
\hat{\phi}_p &= 2\pi\sqrt{\frac{1}{2R_Q}}\sum_{k = 0}^{N}[R]_{p,k}(\hat{a}_k+\hat{a}_k^\dagger).\label{Pi_mode}
\end{flalign}
By imposing the canonical commutation relation for the bosonic operators and
$[\hat{\phi}_p,\hat{n}_m] = i\delta_{pm}$, we obtain the following normalization
condition on the matrices $[R]$ and $[G]$ : 
\begin{equation}
[G][R]^{T} = I.
\label{eq:normalization}
\end{equation}
Using Eq.~(\ref{Q_mode}) and Eq.~(\ref{Pi_mode}) in $\hat{H}_t$, we obtain : 
\begin{flalign}
\hat{H}_t &= \frac{\hbar}{4} \sum_{m,p = 1}^{N+1}
{(\hat{a}_p^\dagger + \hat{a}_p)[G^TC^{-1}G]_{p,m}(\hat{a}_m^\dagger + \hat{a}_m) - 
(\hat{a}_p^\dagger - \hat{a}_p)[R^TL^{-1}R]_{p,m}(\hat{a}_m^\dagger - \hat{a}_m)}.\label{eq:ham1}
\end{flalign}
In order to recover the usual harmonic form (\ref{eq:Htharmo}) of $\hat{H}_t$, 
we firstly impose:
\begin{equation}
[L^{-1}C^{-1}G] = [G\Omega^2],  \\
\label{eq:eom1}
\end{equation}
implying that the columns of $[G]$ contain the right-eigenvectors of $[L^{-1}C^{-1}]$,
$[\Omega]$ being the positive definite diagonal matrix such that $[\Omega^2]$ contains the eigenvalues of $[L^{-1}C^{-1}]$.
Then we note that 
\begin{eqnarray}
[G^T C^{-1}][L^{-1}C^{-1}]&=&[G^T C^{-1}L^{-1}][C^{-1}]\nonumber\\
&=&[\Omega^2 G^T][C^{-1}]=[\Omega^2] [G^TC^{-1}]
\end{eqnarray}
i.e. the rows of $[G^TC^{-1}]$ contain the left-eigenvectors of $[L^{-1}C^{-1}]$, implying that we can take $[G^TC^{-1}G]$ as diagonal.
We have not yet specified the normalization of the columns of $G$. We do so now by imposing
\begin{equation}
[G^TC^{-1}G] = [\Omega]
\end{equation}
From Eq. \ref{eq:normalization} then  follows that $[R^T]=[\Omega^{-1}G^TC^{-1}]$. Using this together with Eq. (\ref{eq:eom1}), we then
derive that also
\begin{equation}
[R^TL^{-1}R] = [\Omega]
\end{equation}
Substitution into Eq. (\ref{eq:ham1}) then yields
\begin{equation}
H_\text{t}=\hbar\sum_{p=1}^{N+1}\omega_p(a_p^\dagger a_p+1/2),
\end{equation}
with $\omega_p=[\Omega]_{pp}$.
Once the $[L^{-1}C^{-1}]$ eigenvalue problem has been numerically solved, we can express the phase across the weak link in terms of
the normal mode amplitudes
\begin{equation}
\phi_k = \pi\sqrt{\frac{2}{R_\text{Q}}}[R]_{0,k},
\end{equation}
so that the final self-consistent equation for $E_\text{J}^*$ is : 
\begin{equation}
E_\text{J}^* =  E_\text{J} \exp(-2\pi^2\sum_{k = 0}^{N}
{\frac{\left[R\right]^2_{0,k}}{R_\text{Q}} \frac{1+2n_k}{2}}) \label{final_Ej}.
\end{equation}
In practice, we determine $E_\text{J}^*$ from the Hamiltonian formalism described
here. Once the value has been determined (which in general depends also on temperature),
it can be inserted in a full ABCD calculation~\cite{Spozar2009microwave}, since the effect of the
capacitive coupling to the output ports is very small in practice.

\subsection{Supplementary Note 6: Phase shift induced by the small Josephson junction}

Now that we have obtained the best harmonic approximation of $\hat{H}$ by solving
(\ref{final_Ej}) self-consistently, we can investigate the effect of the small junction on 
the odd modes with respect to decoupled even modes.
In frequency domain, the equations of motion for the classical phases $\phi_{j,-}$ are given by : 
\begin{equation}
[L^{-1}][\phi_-] =  [C] [\phi_-][\Omega^2],
\end{equation}
with $[L^{-1}]$ and $[C]$ the inductance and capacitance matrices for the odd
modes, and the columns of the matrix $[\phi]$ tabulate the phase configuration for
different frequencies. The even modes form stationary cosine waves  
along the chain: 
\begin{equation}
[\phi_+]_{l,k} = N\cos[k(l+1/2)],
\end{equation}
with $l = 0,1,2...$ the position in the chain and $k$ the wavenumber.
The dispersion relation reads
\begin{equation}
k = 2\arccot\sqrt{\left(\frac{4C}{C_\text{g}}+1\right)
\left(\left[\frac{\omega_\text{p}}{\omega(k)}\right]^2-1\right)} \label{disp_rel},
\end{equation}
with $\omega_\text{p} = 1/\sqrt{L}(C+C_\text{g}/4)$ the plasma frequency of the chain.

In presence of the small junction (treated at the SCHA level), the odd
modes have the same dispersion relation but experience an additional phase shift $\theta$ (we omit in our notation the fact that 
$\theta=\theta_k$ depends implicitely on $k$):
\begin{equation}
[\phi_-]_{l,k} =N\cos[k(l+1/2) - \theta] \label{rel_space}.
\end{equation}
The phase shift is determined from equation of motion that links sites $0$ and $1$
\begin{flalign}
\left(\frac{2}{L^*}+\frac{1}{L}\right)[\phi_-]_{0,k} - \frac{1}{L}[\phi_-]_{1,k} &= \omega^2(C_{\Sigma}[\phi_-]_{0,k} - C[\phi_-]_{1,k}) \label{shift_1},\\
\end{flalign}
which we can rewrite using Eq.\,(\ref{rel_space}) as
\begin{equation}
\cos(k/2-\theta)=\lambda \cos(3k/2-\theta)
\end{equation}
where
\begin{equation}
\lambda = \frac{1 - \omega^2CL}{(1 + \frac{2L}{L_J^*})  -
\omega^2C_\Sigma L} \label{phi1_phi0}.
\end{equation}
In the case where the junction is saturated (either at strong driving power, or
for large thermal fluctuations), we have $E_J^*$ = 0, and we use:
\begin{equation}
\lambda = \frac{1 - \omega^2CL}{1  - \omega^2C_\Sigma L}
\label{phase_shift_csh}.
\end{equation}
Solving for $\theta$, we find
\begin{equation}
\theta=k+\rm{arctan}\left[\frac{(1-\lambda)X}{1+\lambda}\right]\label{th_phase_shift}
\end{equation}
where
\begin{equation}
X =\cot\left(\frac{k}{2}\right)=\sqrt{\left(\frac{4C}{C_\text{g}}+1\right)\left[\left(\frac{\omega_\text{p}}{\omega}\right)^2-1\right]}.
\end{equation}

\newpage
\subsection{Supplementary Note 7: Splitting between odd and even modes}

Now that we have the analytic expression~(\ref{th_phase_shift}) for the phase shift induced 
by the small non-linear junction, we will see how it translates into the splitting 
between odd and even modes. For simplicity, we will assume here that $C_c$ and $C_{c,I}$ are 
big enough so that we can consider the last site $N$ as grounded:
\begin{equation}
k_n(N+1/2) - \theta_n = \pi(n - \frac{1}{2}) \label{splitting_boundary},
\end{equation}    
with $\theta_n$ the phase shift for the mode n, so that : 
\begin{equation}
k_n = k_n^\circ  + \frac{\theta_n}{N+1/2},
\end{equation} 
with $k_n^\circ$ the wave vector of the mode n in the bare chain (corresponding
to the uncoupled even modes in the experiment).
Using the dispersion relation, we find at order $1/N$:
\begin{flalign}
\omega(k_n) &= \omega\left(k_n^\circ  + \frac{\theta_n}{N+1/2}\right)
= \omega(k_n^\circ ) + \frac{\theta_n}{N}\frac{\partial\omega(k)}{\partial k}
\Bigr\rvert_{k = k_n^\circ} + O(N^{-2}) \label{fst_splitting}.
\end{flalign}
We also have for the bare modes:
\begin{flalign}
\omega(k_{n+1}^\circ) = \omega(k_{n}^\circ ) 
+ \frac{\pi}{N}\frac{\partial\omega(k)}{\partial k} \Bigr\rvert_{k = k_n^\circ} + O(N^{-2}) \label{snd_splitting}
\end{flalign}
Using Eq.~(\ref{fst_splitting}) and Eq.~(\ref{snd_splitting}), we obtain the
connection between the relative odd-even splitting $S$ induced by the small junction on the odd modes
and the associated phase shift $\theta_n$ on mode $n$:
\begin{equation}
\theta_n = \pi\frac{\omega(k_n) - \omega(k_n^\circ )}{\omega(k_{n+1}^\circ )
- \omega(k_n^\circ )} = \pi S.
\end{equation}
To make sure that approximating the site $N$ as grounded is valid, we computed numerically
the exact splitting obtained with and without these pads, using a full ABCD
matrix calculation (shown in Supplementary Figure~\ref{fig:ABCDtest} with the parameters of
sample B), and found very little effect of this approximation. In addition, we
find that the theoretical phase shift Eq.~(\ref{th_phase_shift}), valid for an infinite 
chain and shown by the black solid line in Supplementary Figure~\ref{fig:ABCDtest} compares
quantitatively to the ABCD simulations (dots) of the real device.
\begin{figure}[H]
\begin{center}
\includegraphics[width=0.5\textwidth]{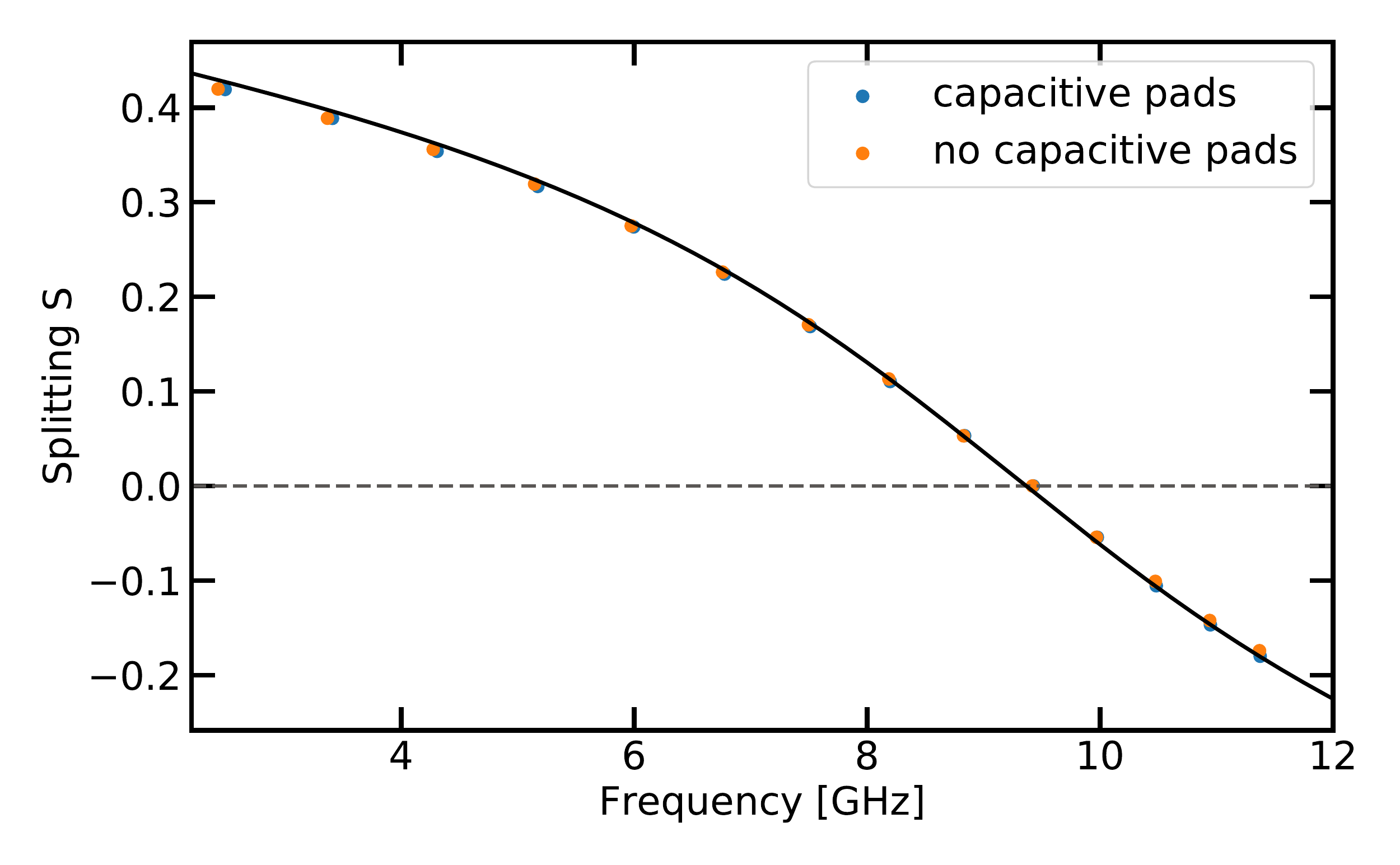}
\caption{{\bf Comparision between the analytical phase shift and the
simulated even-odd splitting.}
The normalized phase shift $\theta_n/\pi$ from formula~(\ref{th_phase_shift}) is in excellent 
agreement with full ABCD simulations of sample B (dots), confirming also a very small effect of 
the coupling pads to the output ports.}
\label{fig:ABCDtest}
\end{center}
\end{figure}
\newpage
In the infinite system, the phase shift $\theta$ becomes a continuous function of frequency
$\omega$. It vanishes at the renormalized frequency
\begin{equation}
\omega_J^*=\frac{1}{\sqrt{L^*(C_\text{J}+C_\text{sh})}}
\end{equation}
of the weak link, as can be seen as follows. When $\theta=0$, Eq.\,(\ref{th_phase_shift})
can be rewritten as 
\begin{equation}
\cot k=\frac{\lambda+1}{(\lambda-1)X}.\label{eq:int}
\end{equation}
From the definition of $X$ follows that $\cot k=(X^2-1)/2X$, and furthermore, that
\begin{equation}
\frac{X^2-1}{2}=2\frac{1-LC\omega^2}{LC_\text{g}\omega^2}-1.
\end{equation}
Using the definition (\ref{phi1_phi0}) of $\lambda$ and that of $C_\Sigma$, 
we reduce Eq.\,(\ref{eq:int}) to
\begin{equation}
\frac{1-LC\omega^2}{LC_\text{g}\omega^2}=\frac{1-LC\omega^2}{\omega^2L[2(C_\text{J}+C_\text{sh})+C_\text{g}]-2L/L^*}
\end{equation}
implying that
\begin{equation}
\omega^2(C_J+C_\text{sh})-1/L^*=0
\end{equation}
and hence
$\omega=1/\sqrt{L^*(C_\text{J}+C_{\text{sh}})}\equiv \omega_\text{J}^*$.

\subsection{Supplementary Note 8: Fitting the experimental splittings}

We present in Supplementary Figure~\ref{fig:SplitFit} the frequency-dependent splittings extracted
from the analysis of the even-odd mode pairs (see Supplementary Figure~\ref{Fig:Fitting}), shown
as dots for our three samples and various temperatures. Each of this data set is
then fitted to the analytical formula~(\ref{th_phase_shift}), $L^*(T)$, 
or equivalently $E_\text{J}^*(T)$ being the fitting parameter.
The range of investigated temperature is restricted below 130 mK, since at too
high temperatures, thermal fluctuations are so strong that the SCHA treatment
breaks down. We find in Supplementary Figure~\ref{fig:SplitFit} that the lineshape of the
splitting is well reproduced by our calculations. The location of the zero of
the splitting also allows to extract the value of the renormalized frequency
$\omega_\text{J}^*$ of the small junction, a key quantity that is discussed in detail
in the main text.
\begin{figure}[H]
\begin{center}
\includegraphics[width=\textwidth]{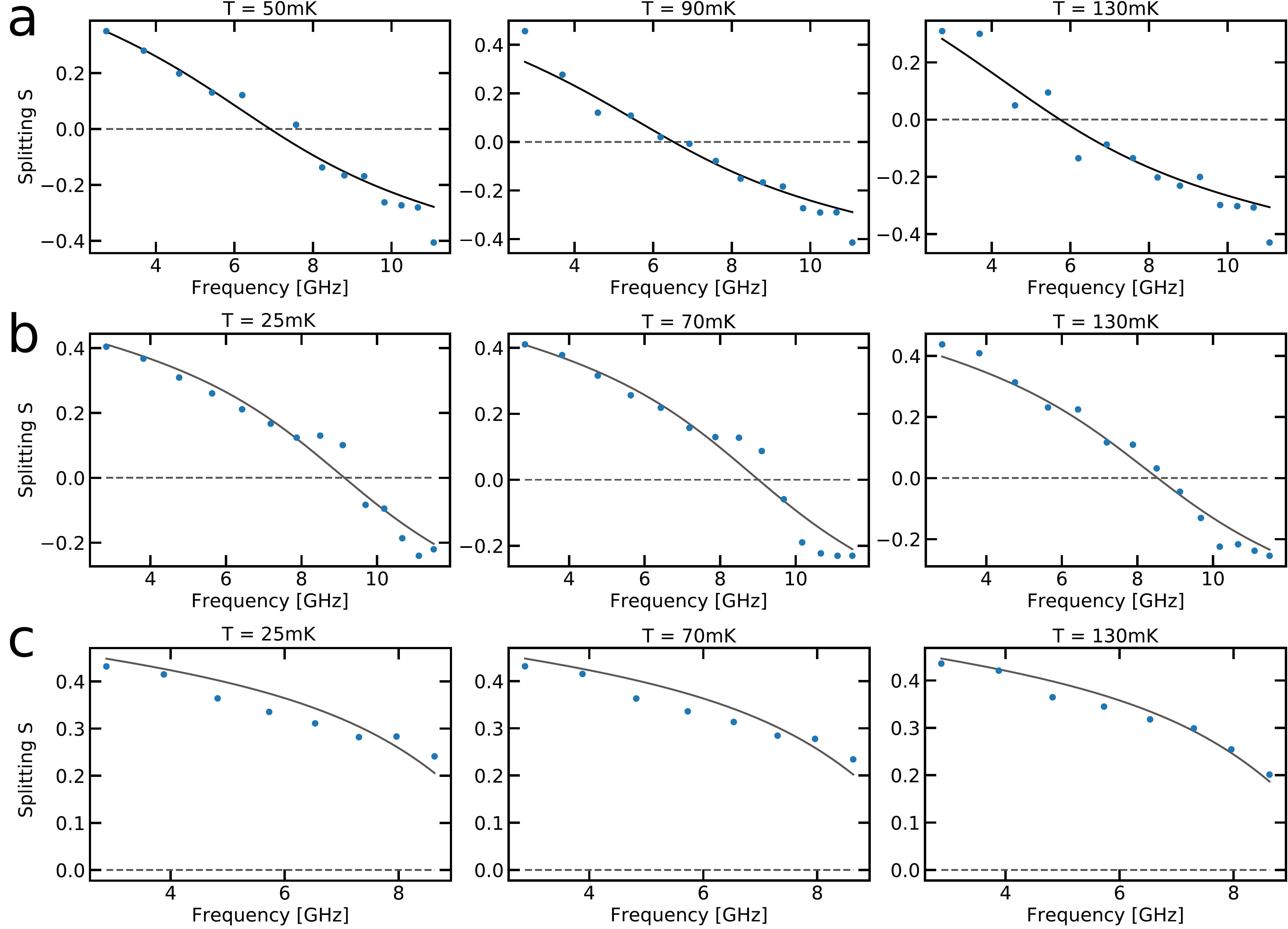}
\caption{ {\bf Analysis of the experimental even-odd splitting.} The
extracted experimental splitting are shown as dots for our three samples 
(\textbf{a} is for sample A, \textbf{b} is for sample 375 and \textbf{c} 
is for sample 450) and various temperatures as indicated.
Formula~(\ref{th_phase_shift}) is fitted (black solid lines), allowing the
extraction of  the renormalized frequency $\omega_\text{J}^*$.}
\label{fig:SplitFit}
\end{center}
\end{figure}

\newpage
\subsection{Supplementary Note 9: Estimation of the shunting capacitance}

To determine a value of the unknown shunting capacitance $C_\text{sh}$, we 
devised an original saturation technique.
At high enough power, the fluctuations across the small junction can be so large
that $E_\text{J}^*$ renormalizes to zero, decoupling effectively the dynamics of the two 
chains, except for the remaining effect of $C_\text{sh}$ and $C_\text{J}$. We can thus use
formula~(\ref{phase_shift_csh}), and since $C_\text{J}$ is known by design, one can
directly infer $C_\text{sh}$ from an analysis of the even-odd splitting at high
power. The evolution of the transmission as a function of power, and the
resulting splittings are shown in Supplementary Figure~\ref{fig:power}.
From that measurement one can infer that $C_\text{sh}$ slightly increases (see
Table I in the main text) when the size of the junction is increased, which is
the expected behavior.
\begin{figure}[H]
\begin{center}
\includegraphics[width=\textwidth]{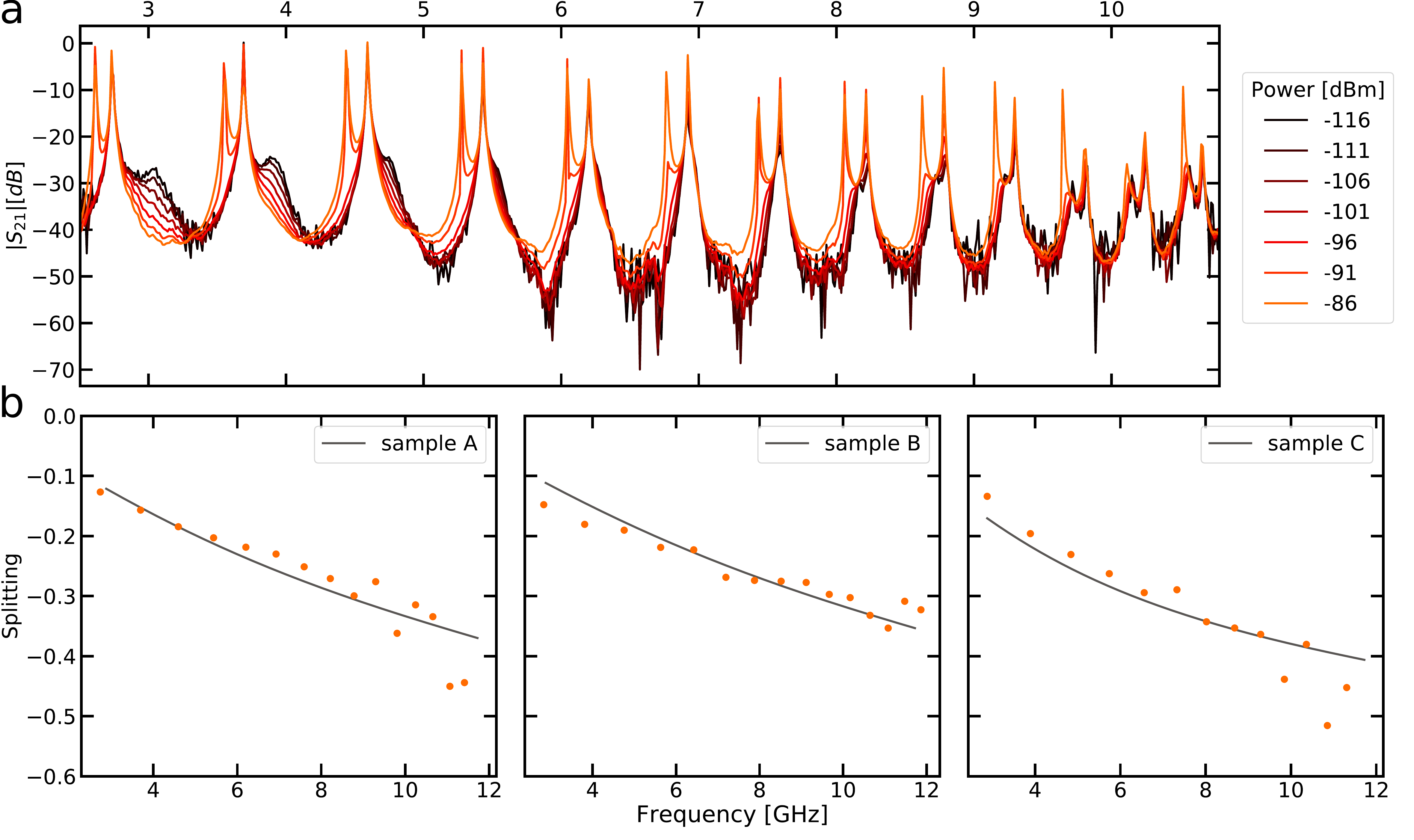}
\caption{{\bf Power scan.} Panel \textbf{a} shows the
transmission $|S_{21}|$ of sample 300 as a function of the frequency for
different power values imposed to the sample. Panel \textbf{b} shows
the extracted splitting for our three samples, fitted from Eq.~(\ref{phase_shift_csh}), 
allowing to extract the shunting capacitance $C_\text{sh}$.}
\label{fig:power}
\end{center}
\end{figure}

\subsection{Supplementary Note 10: Extracting the parameters of the chain}

In this section we discuss how the parameters of the chain are extracted. The chains 
used in our samples are made out of SQUIDs. Consequently, the inductance of the chains
are given by : 
\begin{equation}
L = \frac{L_{\text{J}_\text{ch,min}}}{\sqrt{\cos[2](\Phi_\text{C}/\Phi_0) + d^2
\sin[2](\Phi_\text{C}/\Phi_0)}}
\label{LJch}
\end{equation}
with $\Phi_\text{C}$ the flux in the SQUID loops and $d$ the asymmetry of the SQUID junctions~\cite{SPuertasMartinez:2019gk}.
As we can neglect the effect output port capacitances, the dispersion relation of
the even modes is given by (\ref{disp_rel}), which can be expressed as a
function of $\omega$:
\begin{equation}
\omega(k) = \frac{1}{\sqrt{L(\Phi_\text{C})C}}\sqrt{\frac{1-\cos(ka)}{1-\cos(ka) + \frac{C_\text{g}}{2C}}} \label{disp_rel_2}
\end{equation} 
From Eq.~(\ref{LJch}), the free spectral range (namely the energy difference
between two consecutive modes) is decreasing when $\Phi_\text{C}/\Phi_0$ goes to
$\pi/2$. This behavior is clearly seen in Supplementary Figure~\ref{fig_flux}. One can also notice 
the absence of artifacts around $\Phi_\text{C} = 0$, which means that the chain is homogeneous
and relatively exempt of disorder.
By doing a two-tone spectroscopy at $\Phi_\text{C} = 0$, we can measure precisely the
dispersion of the even modes up to 14GHz. From Eq.~(\ref{disp_rel_2}), we find $C_\text{g}$ and $L$ for 
the three sample, $C$ being known by design. This method allows an in-situ
determination of the chain parameters.

\begin{figure}[H]
\begin{center}
\includegraphics[width=\textwidth]{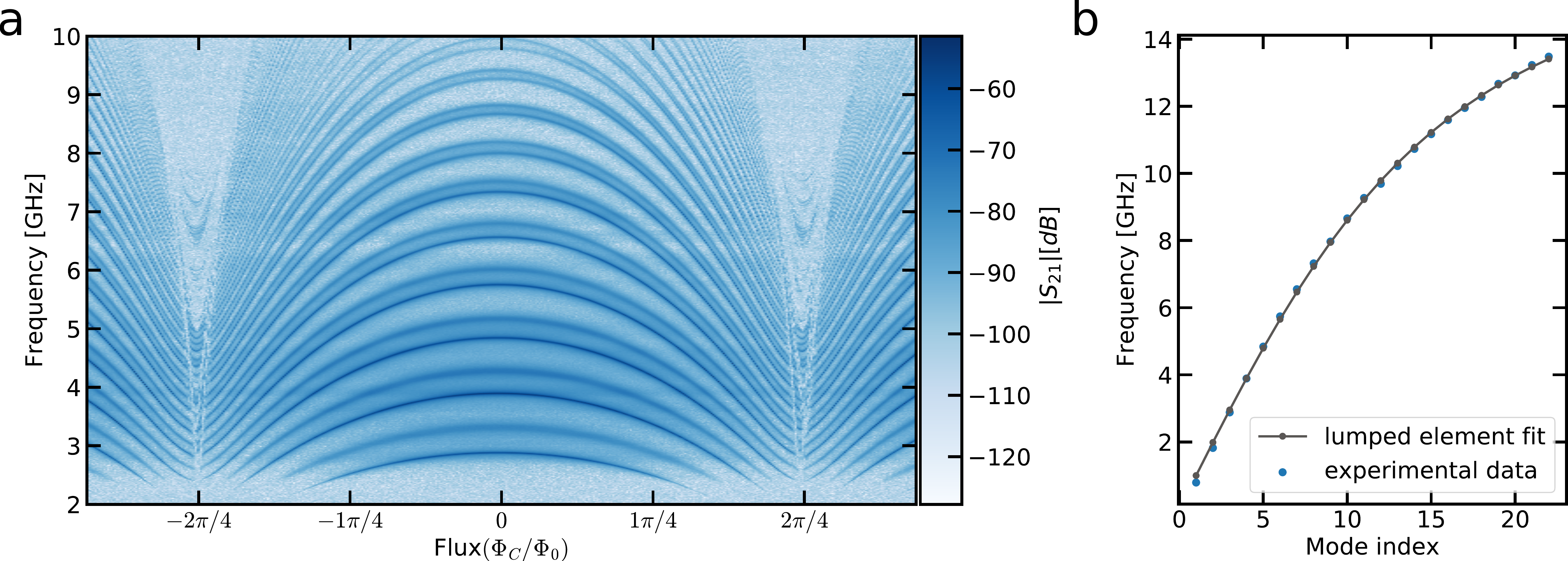}
\caption{{\bf Determination of the chain parameters.} Panel \textbf{a}
shows the transmission $|S_{21}|$ of sample C as a function of the frequency for different flux 
in the SQUIDs of the chain. Panel \textbf{b} shows the dispersion relation for
the even modes extracted at half flux quantum.}
\label{fig_flux}
\end{center}
\end{figure}

\subsection{Supplementary Note 11: Perturbative treatment of the non linearity}

The perturbative treatment is commonly used in circuit-QED whenever one needs to
consider the non-linearity induced by a Josephson junction in a superconducting
circuit~\cite{SAnonymous:2012ek,SWeissl:2015do}. As a first step, the tunnelling energy 
$E_\text{J,bare}(1 - \cos\hat{\phi}_\text{J})$ is approximated by its harmonic
approximation $E_\text{J,bare}\hat{\phi}_\text{J}^2/2$, leading to an effective quadratic
Hamiltonian (without any renormalization)
$\hat{H}^{\text{lin}} = \sum_{k = 0}^M{\hbar\omega^{\text{lin}}_{k}\hat{a}^\dagger_{k}\hat{a}_{k}}$
and a mode decomposition of the phase fluctuating across the weak link
$\hat{\phi}_\text{J} = \sum_{k = 0}^{M}{\phi^{\text{lin}}_k(\hat{a}_k^\dagger +
\hat{a}_k)}$. The non linearity is then reintroduced at quartic level:
\begin{equation}
	\hat{H} \simeq \hat{H}^{\text{lin}} - \frac{E_\text{J}}{24}\hat{\phi}_\text{J}^4.
\end{equation}
This quartic perturbation renormalizes the modes at order $E_\text{J}$:
\begin{flalign}
\omega_{k}^* &= \omega_{k} - [K]_{k,k} + \sum_{j\neq k}{[K]_\text{k,j}}, \\
 K_{k,j} &= E_\text{J}\left(\frac{\pi}{R_\text{Q}}[R]_{0,k}[R]_{0,j}\right)^2.
\end{flalign}
$[K]$ is known as the Kerr matrix~\cite{SKrupko:2018is}.
Using this formalism, one can compute the splitting between odd and even modes
(see Supplementary Figure~\ref{fig:compare}). Fitting with the phase shift
formula~(\ref{th_phase_shift}), we deduce the renormalized Josephson energy from the 
Kerr theory $E_\text{J,Kerr}^*$, which can be compared to the SCHA estimate $E_\text{J,SCHA}^*$ and 
the bare value $E_\text{J,bare}$.
For sample B, we find $E_\text{J,bare} = 5.61$ GHz, $E_\text{J,Kerr} = 4.47$ GHz and 
$E_\text{J,SCHA} = 3.30$ GHz. The renormalization of $E_{J,\text{bare}}$ from the SCHA acquires a clear
non-perturbative character, which the standard Kerr approach is unable to
predict quantitatively. This confirms that our device operates in the many-body
regime, and cannot be described by standard approaches such as
black-box-quantization~\cite{SAnonymous:2012ek}.
\begin{figure}[H]]
\begin{center}
\includegraphics[width=0.5\textwidth]{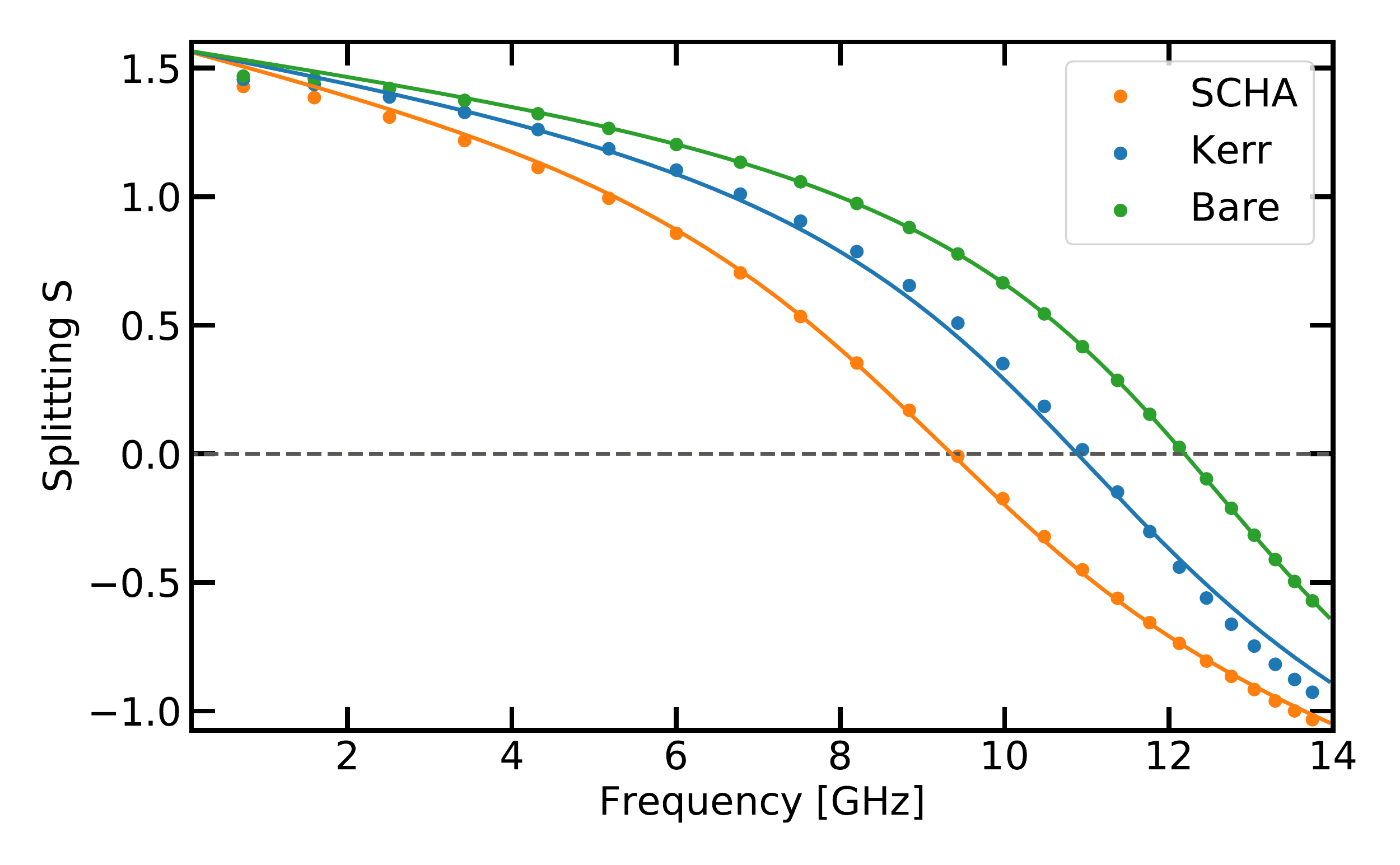}
\caption{{\bf Comparing various quantum approaches of many-body circuits.}
The even-odd splittings for parameters of sample B are obtained from three numerical approaches 
(dots): a bare formalism using the fully linearized Josephson Hamiltonian (green), a Kerr
approach incorporating the quartic correction (blue), and the self-consistent
harmonic approximation taking into account the full cosine form of the potential
(orange). Solid lines are the fits from the phase shift formula~(\ref{th_phase_shift})
allowing to extract the resonance frequency of the junction, and its associated
Josephson energy.}
\label{fig:compare}
\end{center}
\end{figure}

%

\end{document}